\documentclass{aa}
\usepackage{natbib,psfig}

\def\mathnew{\mathsurround=0pt}
\def\simov#1#2{\lower .5pt\vbox{\baselineskip0pt \lineskip-.5pt
\ialign{$\mathnew#1\hfil##\hfil$\crcr#2\crcr\sim\crcr}}}

\def\MeV{Me\kern-0.11em V}
\def\keV{ke\kern-0.11em V}

\begin{document}

\title{Galaxy cluster searches based on photometric redshifts \\ in
  the four CFHTLS Wide fields \thanks{Based on observations obtained with
    MegaPrime/MegaCam, a joint project of CFHT and CEA/DAPNIA, at the
    Canada-France-Hawaii Telescope (CFHT) which is operated by the
    National Research Council (NRC) of Canada, the Institut National
    des Sciences de l'Univers of the Centre National de la Recherche
    Scientifique (CNRS) of France, and the University of Hawaii. This
    work is based on data products produced at TERAPIX and the
    Canadian Astronomy Data Centre as part of the Canada-France-Hawaii
    Telescope Legacy Survey, a collaborative project of the NRC and
    CNRS.}}

\offprints{F. Durret  \email{durret@iap.fr}}

\author{ F.~Durret\inst{1,2}
\and
C. Adami\inst{3}
\and
A.~Cappi\inst{4,5}
\and
S.~Maurogordato\inst{5}
\and
I.~M\'arquez\inst{6}
\and
O.~Ilbert\inst{3}
\and
J.~Coupon\inst{1,2,7}
\and
S.~Arnouts\inst{8}
\and
C.~Benoist\inst{5}
\and
J.~Blaizot\inst{9}
\and
T.M.~Edorh\inst{10}
\and
B.~Garilli\inst{3,11}
\and
L.~Guennou\inst{3}
\and
V.~Le~Brun\inst{3}
\and
O.~Le~F\`evre\inst{3}
\and
A.~Mazure\inst{3}
\and
H.J.~McCracken\inst{1,2}
\and
Y.~Mellier\inst{1,2}
\and
C.~Mezrag\inst{12}
\and
E.~Slezak\inst{5}
\and
L.~Tresse\inst{3}
\and
M.P.~Ulmer\inst{13}
}

\institute{
UPMC Universit\'e Paris 06, UMR~7095, Institut d'Astrophysique de Paris,
98bis Bd Arago, F-75014, Paris, France
\and
CNRS, UMR~7095, Institut d'Astrophysique de Paris, F-75014, Paris, France
\and
LAM, OAMP, P\^ole de l'Etoile Site Ch\^ateau-Gombert, 38 rue Fr\'ed\'eric Juliot-Curie,  13388 Marseille Cedex 13, France
\and
INAF - Osservatorio Astronomico di Bologna, via Ranzani 1, 40127 Bologna, Italy
\and
OCA, Cassiop\'ee, Boulevard de l'Observatoire, B.P. 4229, F-06304 NICE Cedex 4, France
\and
Instituto de Astrof\'\i sica de Andaluc\'\i a (CSIC), Apdo. 3004, 18080 Granada, Spain 
\and
Astronomical Institute, Graduate School of Science, Tohoku University
Sendai 980-8578, Japan
\and
Canada-France-Hawaii Telescope Corporation, Kamuela, HI-96743, USA
\and
CRAL (UMR 5574), Universit\'e Claude Bernard Lyon 1 (UCBL),
 Ecole Normale Sup\'erieure de Lyon (ENS-L), and Centre National de la Recherche
 Scientifique (CNRS), Saint-Genis Laval F-69230, France
\and
Universit\'e Paris 07 Denis Diderot, F-75205 Paris cedex, France
\and
INAF IASF - Milano, via Bassini 15, 20133 Milano, Italy
\and
ENS-Cachan, 61, avenue du pr\'esident Wilson, F-94235 Cachan cedex, France
\and
Department of Physics $\&$ Astronomy, Northwestern University, Evanston, IL 60208-2900, USA
}

\date{Accepted . Received ; Draft printed: \today}

\authorrunning{Durret et al.}

\titlerunning{Galaxy cluster searches in the CFHTLS Wide fields}

\abstract
{Cosmological parameters can be constrained by counting clusters of
  galaxies as a function of mass and redshift, and by considering regions
  of the sky sampled as deeply and as homogeneously as possible.  }
{Several methods for detecting clusters in large imaging surveys have
  been developed, among which  the one used here, which is based on 
  detecting of structures. This method was first applied to the Canada
  France Hawaii Telescope Legacy Survey (CFHTLS) Deep~1 field by
  Mazure et al. (2007), then to all the Deep and Wide CFHTLS fields
  available in the T0004 data release by Adami et al. (2010). The
  validity of the cluster detection rate was estimated by applying the
  same procedure to galaxies from the Millennium simulation. Here we use
  the same method to analyse the full CFHTLS Wide survey, based on
  the T0006 data release.}
{Our method is based on the photometric redshifts computed with Le
  Phare for all the galaxies detected in the Wide fields, limited to
  magnitudes $z' \leq 22.5$. We constructed galaxy density maps in
  photometric redshift bins of 0.1 based on an adaptive kernel
  technique, detected structures with SExtractor at various detection
  levels, and built cluster catalogues by applying a minimal spanning
  tree algorithm.  }
{In a total area of 154~deg$^2$, we have detected 4061 candidate
  clusters at 3$\sigma$ or above (6802 at 2$\sigma$ and above), in the
  redshift range $0.1\leq z \leq 1.15$, with estimated mean masses between
  1.3 10$^{14}$ and 12.6 10$^{14}$ M$_\odot$. This catalogue of
  candidate clusters will be available online via VizieR.  We compare
  our detections with those made in various CFHTLS analyses with other
  methods.  By stacking a subsample of clusters, we show that this
  subsample has typical cluster characteristics (colour-magnitude
  relation, galaxy luminosity function).  We also confirm that the
  cluster-cluster correlation function is comparable to the one obtained
  for other cluster surveys and analyse large-scale filamentary galaxy
  distributions.}
{We have increased the number of known optical high-redshift cluster
  candidates by a large factor, an important step towards obtaining
  reliable cluster counts to measure cosmological parameters. The
  clusters that we detect behave as expected if they are located at
  the intersection of filaments by which they are fed. }


\maketitle

\section{Introduction}\label{sec:intro}

The cluster count technique (e.g. Gioia et al. 1990) allows putting
strong constraints on cosmological parameters, but requires catalogues
with large numbers of clusters at high redshift (z$\geq$1) and in
extended fields of view (several tens of square degrees). Many large
surveys have been done in the past ten years. One of their aims was
to obtain large catalogues of galaxy clusters, among these the
Canada France Hawaii Telescope Legacy Survey
(CFHTLS)\footnote{http://www.cfht.hawaii.edu/Science/CFHLS/}.  The
CFHTLS observations are carried out in five filters ($u^*,\ g',\ r',\
i'$ or $y$, and $z'$) providing catalogues of sources that are 80\%
complete up to $i_{AB}$=24.0 for the CFHTLS Wide (see Goranova et al.
2009).  The CFHTLS survey encloses a sample of about $20\ 10^6$
galaxies in a volume of about 1~Gpc$^3$, with a median redshift of z$
\sim 0.60$ within a limiting magnitude $i'\leq 23$ for the Wide survey
(see Table~6 in Coupon et al. 2009). Romer et al. (2001) 
estimate that about 20,000 clusters with temperatures $kT>2$~keV and
redshifts $z\leq 1.5$ were expected in 800~deg$^2$ assuming a standard
cosmological model. The corresponding number of clusters expected in
the 154~deg$^2$ covered by the CFHTLS (not including the Deep survey)
would then be 3850, consistent with the number of cluster candidates
we found. We are indeed considering comparable mass ranges such
as Romer et al. (2001): our 3$\sigma$ detections correspond to a
minimal mass of $1.3\ 10^{13}$~M$_\odot$ and to a mean mass of $1.8\
10^{14}$~M$_\odot$ (see Table~\ref{tab:minmass}), while the scaling
relation shown by Juett et al. (2010) indicates that clusters with
kT$>2$~keV have masses of M$_{500}$E(z)$>10^{14}$~M$_\odot$.

Early searches for clusters of galaxies in the CFHTLS were performed
by Olsen et al. (2007 and 2008) and Grove et al. (2009), based on a
matched filter detection algorithm applied to the Deep fields. An
improvement of this technique has recently been developed by
Milkeraitis et al. (2010) and also applied to the CFHTLS Deep
fields. Lensing techniques were employed to detect massive structures
(i.e. with masses over $10^{13}$~M$_\odot$) in the CFHTLS
(e.g. Cabanac et al. 2007, Gavazzi $\&$ Soucail 2007, Berg\'e et
al. 2008). Other cluster studies based on the CFHTLS data (e.g. the
CFHTLS-CARS survey: Erben et al. 2009, Hildebrandt et al. 2009) and
based in part on the red sequence in the colour magnitude diagram have
also been developed.  Bielby et al. (2010) have recently identified
high-redshift ($z>1.1$) groups and clusters in the Deep~1 and Deep~4
fields, by combining CFHTLS optical data with the WIRCam Deep Survey
(WIRDS) in the infrared and XMM-Newton data. Ascaso et al. (2010)
applied a Bayesian cluster finder to detect galaxy clusters in several
surveys including the CFHTLS, and have detected 90\% of the clusters
found by Olsen et al. (2007) and Adami et al. (2010, hereafter A10).

We have developed a new method to search for clusters in large
multiband imaging surveys and applied it to the CFHTLS.  The first
results were presented by Mazure et al. (2007) for the
CFHTLS Deep~1 field. More extensive results were later obtained by A10
for the CFHTLS Deep~2, Deep~3 and Deep~4 fields, as well as for the
CFHTLS W1, W3, and W4, as available in 2008 (T0004 data
release). A10 also applied the same method to the Millennium
simulation, in order to assess the validity of the method and
results. No other data have been obtained for the Deep fields since
then, so we will not discuss these fields any further.  On the other
hand, the CFHTLS is now complete and the coverage of the Wide fields
has become much broader in the T0006 data release, leading us to
reanalyse all the Wide fields in a consistent way.

Thanjavur et al. (2009) developed another method of detecting galaxy
clusters named K2, which they applied to the CFHTLS Wide fields (T0005
data release).  This method is based on the red sequence, and it
detects cluster enhancements in both colours and position. Since these
authors have kindly made their catalogues available to us, we will be
able to compare our list of cluster candidates directly with theirs
(see Section~4).

Van Breukelen \& Clewley (2009) have developed yet another algorithm,
named 2TecX, to search for high-redshift clusters in optical/infrared
imaging surveys.  This method is based on photometric redshifts
estimated from the full redshift probability function and on
identifying cluster candidates through cross-checking two different
selection techniques (adaptations of the Voronoi tesselations and of
the friends-of-friends method). This method is not all that different
from ours, and it would be very interesting to apply it to the CFHTLS
Wide data, to compare the cluster candidates obtained by their method
and by ours. A comparable method has been applied to the SDSS Stripe
82 by Geach et al. (2011) to search for clusters up to z$\sim 0.6$.

The paper is organized as follows. The method of searching for clusters
is described in Section~\ref{sec:method}. Results for cluster
candidates are described in Section~\ref{sec:results}: numbers,
spatial distribution, redshift distribution. Full tables of our
cluster detections are available electronically.  In
Section~\ref{sec:comparison} we compare our cluster candidates in the
W1 field to those found by other authors. The angular correlation
function is discussed in Section~\ref{sec:acf}. We discuss the
cosmological implications of our work in Section~\ref{sec:lss} and
conclude in Section~\ref{sec:concl}.

In this paper we assume H$_0$ = 70 km s$^{-1}$ Mpc$^{-1}$, $\Omega
_m$=0.3, and $\Omega _{\Lambda}$=0.7. All coordinates are given at the J2000
equinox and magnitudes are in the AB system.

\section{The method to search for clusters in the CFHTLS}
\label{sec:method}

We adopt here the same method as in M07 and A10, where a full
description is given.  As mentioned above, we have redone the analysis
for the four full Wide fields, for which the data are now much more
extensive and of better quality than those available in the T0004 data
release on which the A10 paper was based: the total surface covered by
the CFHTLS data is now about 154~deg$^2$, which is about 4.4 times more
extended than the 35~deg$^2$ covered by the T0004 data release, the
photometric zero points were homogenized throughout the entire Wide
survey, and the spectroscopic sample was larger, thus allowing more
accurate computations of the photometric redshifts.

Our approach is based on photometric redshifts computed for all the
galaxies extracted in each field (Coupon et al. 2009), taking the full
colour information into account. We divide the galaxy catalogues in
slices of 0.1 in redshift, each slice overlapping the previous one by
0.05, and build galaxy density maps for each redshift
slice. Structures in these density maps are detected with the
SExtractor software (Bertin \& Arnouts 1996) in the different redshift
bins at various significance levels. We then analyse possible
substructuring by applying a minimal spanning tree algorithm to our
data. In order to assess the validity of this method, the same
treatment was applied by A10 to the Millennium simulation, and we
refer the reader to this paper.

\subsection{Photometric redshifts}

\begin{figure}
\centering
\mbox{\psfig{figure=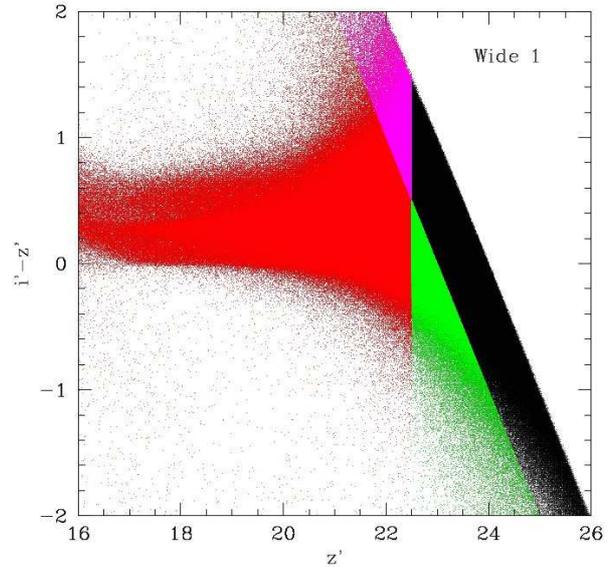,width=8cm}}
\caption[]{$i'-z'$ versus $z'$ colour magnitude diagram for galaxies
  in the Wide~1 field. Black points represent the entire galaxy
  sample. Red+magenta dots are the galaxies corresponding to our
  present selection ($z'\leq 22.5$ and $0.1\leq zphot \leq 1.15$).
  Magenta dots are objects included with our present selection that
  would not have been included with the $i'\leq 23$ selection of
  A10. Green dots are objects included by the A10 selection but not by
  the present selection. The limit to the black points is due to
    an initial magnitude cut in the initial catalogue of galaxy
    photometric redshifts at $i'<24$ or $y< 24$ (depending on the i
    filter available).}
\label{fig:imoinsz_z_all}
\end{figure}

\begin{table*}
  \caption{Numbers of points corresponding to the three different selections 
    illustrated in Fig.~\ref{fig:imoinsz_z_all} for the four Wide fields,
    and fractions of ``gained'' and ``lost'' points due to the change in 
    selection criterium.}
\begin{center}
\begin{tabular}{rrrrr}
\hline
\hline
Field & Wide~1 & Wide~2 & Wide~3 & Wide~4 \\
\hline
Surface (deg$^2$) & 65.25 & 22.28 & 45.54 & 20.88 \\
\hline
B=Black & 5,346,671 & 2,100,453 & 3,858,738 & 2,040,682 \\
RM=Red+magenta & 2,105,404 & 945,090 &  1,517,133 & 977,875 \\
M=Magenta &  98,954 & 42,529 &  77,374 & 36,181 \\
G=Green & 397,921 & 135,148 & 274,601 & 145,703 \\
\hline
M/RM & 0.047 & 0.045 & 0.051 & 0.037  \\
G/RM & 0.189 & 0.143 & 0.181 & 0.149 \\
\hline
\end{tabular}
\end{center}
\label{tab:selection}
\end{table*}

The Wide fields are mosaics of $1\times 1$~deg$^2$ Megacam fields
observed in the $u^*,\ g',\ r',\ i'$, and $z'$ bands. 
Our approach is based on photometric redshifts, which can be estimated
with good precision up to z$\sim$1.2 (Mellier et al. 2008, Coupon et
al. 2009).  Photometric redshifts were computed for all the objects in
the CFHTLS galaxy catalogues of the data release T0006 with the Le
Phare software developed by S.~Arnouts and O.~Ilbert (Ilbert et
al. 2006; also see http://www.ifa.hawaii.edu/\~ ilbert/these.pdf.gz,
pages 50 and 142). They were optimized with spectroscopic redshifts
from the literature, including the VVDS (e.g. Le~F\`evre et
al. 2005). The resulting statistical errors on the photometric
redshifts ($\sigma _{\Delta z/(1+z_S)}$) are given in Coupon
et al. (2009).  For example in the W1 field, they continuously
increase (between $i'=20.5$ and $i'=24$) from 0.025 to 0.053. At our
limiting magnitude of $z'=22.5$ (roughly corresponding to $i'=23$),
the statistical error on the redshift is 0.043. Extensive tests of
photometric redshift software performed by Hildebrandt et al. (2010)
have demonstrated the excellent results of the Le~Phare technique.

To avoid incompleteness effects and strong systematic biases
in photometric redshift computations, the galaxy catalogues were
limited to $z'\leq 22.5$. Galaxy magnitude histograms show that this
limit is located $\sim 0.5$ magnitude below the value where
incompleteness begins to show, therefore our analysis should not
suffer from incompleteness effects.  In our previous analysis of the
Wide fields, we  limited our catalogues to $i'\leq 23$, but we could
not apply this condition in the present work because the $i'$ filter
had to be replaced by a new similar (but not identical) filter $y$
between data releases T0004 and T0006, so a selection in the $i'$ band
would no longer be homogeneous. The T0004 data in the $i'$ band
have not been discarded but merged into the T0006 data.

To test the consistency of the present $z'\leq 22.5$ selection with
the previous $i'\leq 23$ selection, we show in
Fig.~\ref{fig:imoinsz_z_all} the $i'-z'$ versus $z'$ colour magnitude
diagram for galaxies in the Wide~1 field. The plots for the three
other Wide fields are similar. This plot illustrates that by
considering a sample with $z'\leq 22.5$ we are including the magenta
points (which would not have included by the $y\leq 23$ condition) and
losing the green points (those with $y\leq 23$ but $z'>22.5$). The
numbers of points corresponding to each criterium are given in
Table~\ref{tab:selection} for the four Wide fields.  By applying the
$z'\leq 22.5$ condition, we therefore increase the number of galaxies
by 4\% to 5\% and decrease it by 14\% to 19\% relative to the $y\leq
23$ condition.

We selected galaxies with photometric redshifts included in the range
$0.1 \leq z \leq 1.15$. For each CFHTLS Wide field, we give in
Table~\ref{tab:selection} the numbers of galaxies considered
(i.e. with magnitudes $z'\leq 22.5$) and the approximate total surface
covered. Owing to masked areas, the effective total surface covered may
be reduced by as much as 10\% relatively to these values. As a
comparison, the surfaces covered by the T0004 Data Release and
analysed by A10 were approximately 19, 5, and 11~deg$^2$ for the W1, W3,
and W4 fields respectively. W2 was not available.

\begin{figure}
\centering
\mbox{\psfig{figure=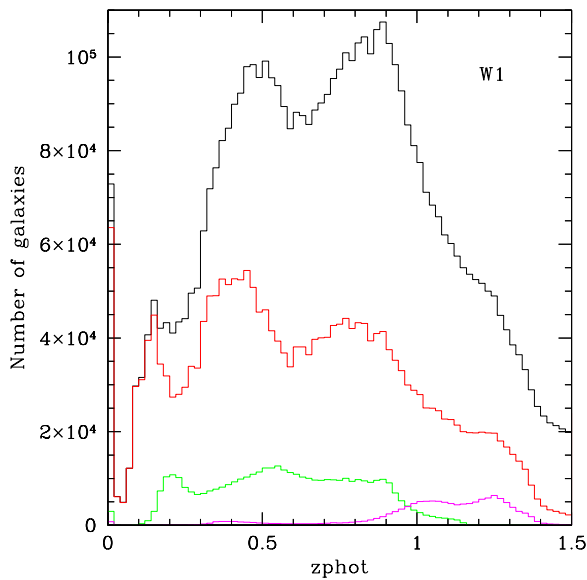,width=8cm}}
\caption[]{Photometric redshift histograms of the galaxies in the
  Wide~1 field for the various magnitude selections described
  above. The colours of the lines are the same as the colours of the
  points in Fig.~\ref{fig:imoinsz_z_all}.}
\label{fig:histozgal} 
\end{figure}

To see how the different magnitude selection criteria (i.e. $z'\leq
22.5$ versus $y\leq 23$) can affect the redshift distributions of the
detected clusters, we show in Fig.~\ref{fig:histozgal} the galaxy
redshift distributions corresponding to the various magnitude
selections described above for the Wide~1 field. We see that with our
new magnitude selection criterium we are missing
low/intermediate-redshift objects but gaining high-redshift red
objects. We do not show the photometric redshift histograms for the
three other fields since they are similar. That three peaks are seen
at redshifts of about 0.15, 0.5, and 0.9 in all four Wide fields is
obviously due to degeneracies in the photometric redshift
computations.

\subsection{Density maps}

For each Wide field, galaxy catalogues were built in running slices of
0.1 in redshift, shifted by 0.05 (i.e. the first slice covers
redshifts 0.1 to 0.2, the second 0.15 to 0.25, etc.). We assumed
the most likely photometric redshift (i.e. the one with the minimum
  $\chi ^2$ value) for each object in order to assign it to a
redshift slice.

As already mentioned in A10, the 0.1 redshift width of the studied
slices is the best compromise between the redshift resolution and the
possible dilution of the density signal due to photometric redshift
uncertainties. Assuming the worst possible photometric redshift
statistical error (for $z'=22.5$, see Coupon et al. 2009) leads to a
1$\sigma$ error of 0.09 at z=1.15 (the upper limit in redshift for the
wide field analyses).

Since the catalogues thus obtained were too large to allow direct
computation of density maps, we cut them into square regions of
$0.9\times 0.9$~deg$^2$, with an overlap of 0.1~deg in both
directions.

Density maps were then computed for each subcatalogue in each redshift
slice, based on an adaptative kernel technique described in M07, with
1000 bootstrap resamplings of the maps to have the correct background
level. The pixel size was 0.54~arcmin and the highest redshift slice
was $1.05-1.15$.

The SExtractor software was then applied to the galaxy density maps to
detect structures at pre-defined significance levels (hereafter called
S/N) of $2\sigma_S$, $3\sigma_S$, $4\sigma_S$, $5\sigma_S$,
$6\sigma_S$, and $9\sigma_S$ (where $\sigma_S$ is the SExtractor
detection threshold). The $9\sigma_S$ threshold was not considered in
the A10 analysis, but we include it here to have as much information
as possible on the possible presence of very massive clusters. The
$2\sigma_S$ threshold is obviously low, but as shown in
Section~4.2.1. we redetect almost 40\% of the Thanjavur et al. (2009)
clusters at $2\sigma_S$, so we can expect that about 40\% of the
clusters that we detect at a $2\sigma_S$ level are real.

The structures were then assembled in larger structures (called
$detections$ in the following) using a friends-of-friends algorithm
(see Adami \& Mazure 1999). In A10, two $detections$ with centres
distant by less than 2.7~arcmin (the typical error on cluster
positions, as estimated by A10) were merged into a single one that was
assigned the redshift of the highest S/N $detection$. Here we chose a
more conservative value of 3.0~arcmin and decided not to merge
$detections$ within 3.0~arcmin into a single one if their photometric
redshifts differed by more than 0.2. This choice was made to avoid
losing clusters that could be more or less aligned along the line of
sight but located at very different redshifts. The difference on the
number of candidate clusters in the Wide~1 field when choosing 2.7 or
3~arcmin is 10.5\%.

The uncertainties on the positions of the cluster centres estimated in
A10 from the comparison with the Millennium simulations were computed
in the following way. We selected all the Millennium haloes present in
a detection ellipse and more massive than $10^{12}$~M$_\odot$.  The
error on the position of the ellipse centre was assumed to be the
distance from the closest Millennium halo. This process
naturally underestimates the uncertainty on the centre position, as
the Millennium simulation always provides more than one halo more
massive than $10^{12}$~M$_\odot$ per detection ellipse.

For a given detection, if we had instead computed the mean difference
between the ellipse centre and all the Millennium halo centres, the
uncertainty on our ellipse position would have been $2.9 \pm
2.0$~arcmin, close to the predicted ellipse position uncertainty when
comparing our results with other observed catalogs.  We thus obtained
catalogues of galaxy cluster candidates in the various CFHTLS Wide
fields with a specified significance level.

\subsection{Previous detection rate assessments}

To assess our detection levels, the same method was applied to a
modified version of the Millennium numerical simulation (e.g. Springel
et al. 2005, http://www.mpa-garching.mpg.de/galform/virgo/millennium/
), as described by A10. We refer to this paper (and in particular to
Fig.~5 for the detection success rate and to Fig.~7 for the catalogue
purity) for a full discussion, and only summarize a few points below.

For the Wide survey, only Millennium haloes more massive than 7.5
10$^{13}$ M$_\odot$, can be detected with a success rate over
$\sim$20$\%$. For lower mass systems, the detection rates become quite
low at z$\geq$0.6-0.7.  False detection rates were basically found to
be zero for S/N$\geq$4 and remain small for S/N$\leq$3 and z$\leq$0.8.
The typical uncertainty on the candidate cluster coordinates was
typically $\sim$2.7~arcmin (see A10, section 4.2) and the redshift
uncertainty was smaller than 0.2.

\begin{table}
\caption{Relation between the SExtractor detection threshold and the minimal
  and mean (over all the associated Millennium haloes) cluster masses.}
\begin{center}
\begin{tabular}{rrr}
\hline
\hline
S/N & Minimal mass & Mean mass   \\
($\sigma _S$)      & (M$_\odot$) & (M$_\odot$) \\
\hline
6 & 5.5 10$^{13}$ & 12.6 10$^{14}$  \\
5 & 3.5 10$^{13}$ & 1.3 10$^{14}$   \\
4 & 3.3 10$^{13}$ & 1.8 10$^{14}$   \\
3 & 1.3 10$^{13}$ & 1.8 10$^{14}$   \\
2 & 1.0 10$^{13}$ & 1.3 10$^{14}$   \\
\hline
\end{tabular}
\end{center}
\label{tab:minmass}
\end{table}

A minimal mass based on the photometry (and on the Millennium
simulation halo masses) can be given for each cluster, assuming that
the detection threshold at which the cluster is detected is a rough
estimate of its richness. For each of our detections we have several
Millennium haloes, and the minimal mass for this detection is taken to
be that of the Millennium halo of lowest mass.  The numbers were given
by A10 (Table~2), but we repeat them in Table~\ref{tab:minmass} for
completeness.

As shown in A10, haloes with a total mass lower than 5~10$^{14}$
M$_\odot$ are not strongly substructured while more massive
$detections$ can be strongly substructured. Because the Millennium
simulation covers only an area corresponding to 1~deg$^2$, it includes
no cluster corresponding to a 9$\sigma$ detection in our study, so we
cannot give the corresponding masses in
Table~\ref{tab:minmass}. Because of the mentioned limits of the
Millennium simulation, we cannot estimate the level of real detections
at $z>$0.7 or for masses higher than $2\ 10^{14}$M$_\odot$.

\section{Results: spatial and redshift distributions of the
  $detections$ }
\label{sec:results}

\subsection{$Detection$ counts}

\begin{table}
\caption{Number N of candidate clusters detected in the four Wide fields
at various significance levels.}
\begin{center}
\begin{tabular}{rrrrr}
\hline
\hline
S/N      & W1 & W2 & W3 & W4 \\
$\sigma _S$   &        &        &        &        \\
\hline
9     &   70  &  24 &  71 &  16 \\
6     &  218  &  75 & 178 &  47 \\
5     &  204  &  55 & 173 &  65 \\
4     &  442  & 125 & 222 & 112 \\
3     &  901  & 230 & 441 & 211 \\
2     & 1412  & 410 & 734 & 366 \\
\hline
N(2-9 $\sigma_s$) & 3247  & 919 & 1819 & 817 \\
N(3-6 $\sigma_s$) & 1835  & 509 & 1085 & 451 \\
N(3-9 $\sigma_s$) & 1905  & 533 & 1156 & 467 \\
\hline
Surface S (deg$^2$) & 65.25 & 22.28 & 45.54 & 20.88 \\
\hline
N(2$\sigma_s$-9$\sigma_s$)/S & 49.8 & 41.3 & 39.9 & 39.1 \\
N(3$\sigma_s$-9$\sigma_s$)/S & 28.1 & 22.8 & 23.8  & 21.6 \\
\hline
\end{tabular}
\end{center}
\label{tab:detections}
\end{table}

The full lists of $detections$ with their coordinates, redshift, and
S/N for the four CFHTLS Wide fields will be available electronically
from the VizieR interface of the Simbad
database\footnote{http://vizier.u-strasbg.fr/viz-bin/VizieR}.  The
numbers of candidate clusters are given in Table~\ref{tab:detections}
for each field and significance level. Altogether we detect 6802
candidate clusters in a total surface of about 154 deg$^2$ (at
confidence levels between 2$\sigma$ and $9\sigma$). The number of
clusters detected per square degree is between 39 and 50. If we only
take clusters detected at a significance level of at least 3$\sigma
_S$ into account, the total number becomes 4061, and the numbers per
square degree range between 21 and 28.

Thanjavur et al. (2009) searched for clusters in the W1 field, based
on the T0005 CFHTLS data release. They found 6144 galaxy clusters, of
which 239 are rich clusters, and thus detected about 35
clusters/deg$^2$. This number is intermediate between the densities of
clusters detected at 2$\sigma_s$ and 3$\sigma_s$.  Since about 55\%
of the clusters that we have detected at 2$\sigma_s$ are likely to be
real (see Section~4.1), we can say that the cluster densities that we
give in Table~\ref{tab:detections} agree with the number given by
Thanjavur et al. (2009). These authors have also sent us their cluster
catalogues in the three other Wide fields, and we will make a detailed
comparison of our numbers with theirs in Sect.~4.2.1.

\subsection{Spatial distributions of the $detections$}

\begin{figure*}
\centering
\mbox{\psfig{figure=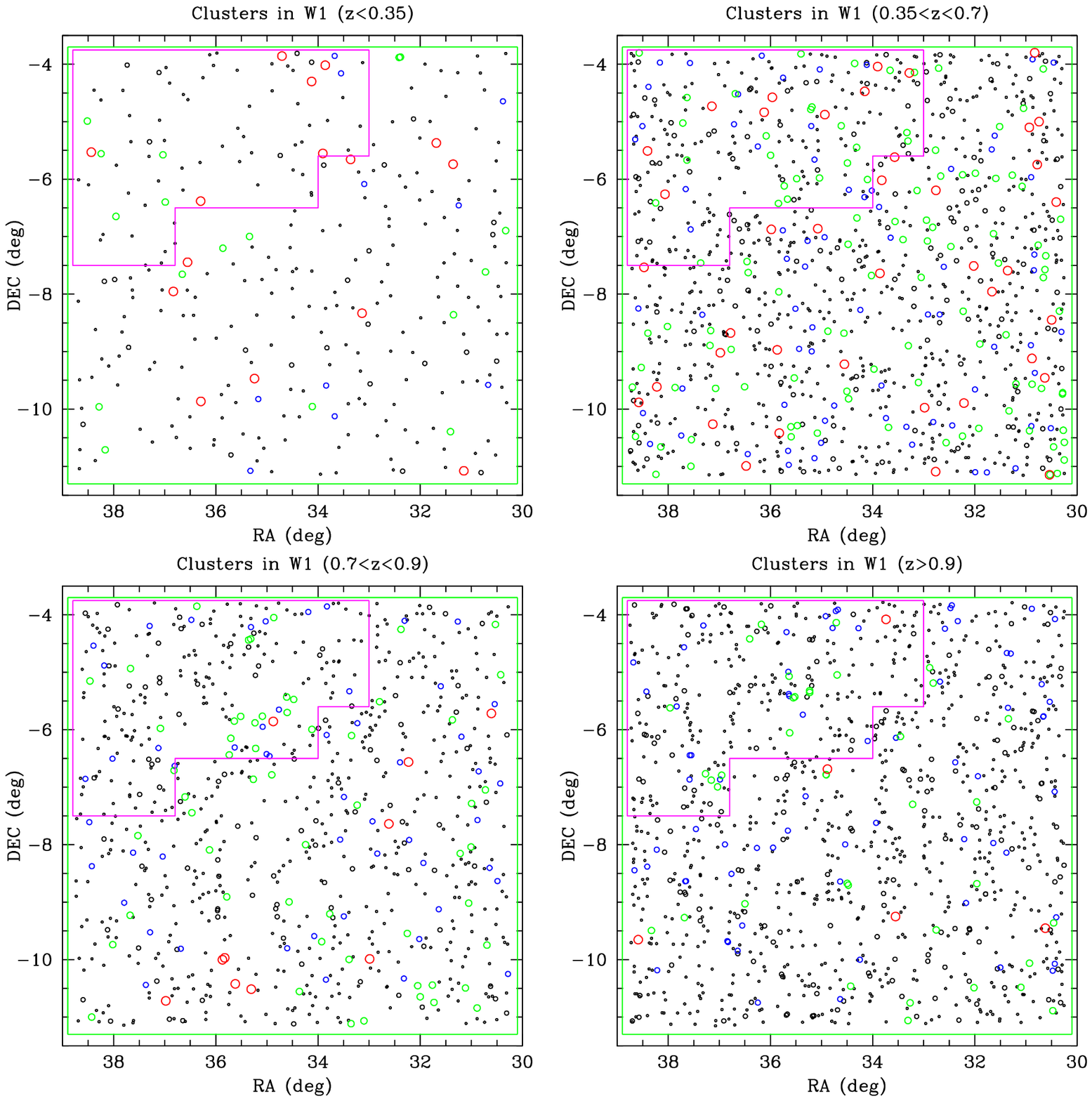,width=16.cm,clip=true}}
\caption[]{Spatial distribution of our $detections$ in the W1 field,
  divided in four photometric redshift bins for the sake of clarity:
  top left z$<$0.35; top right: $0.35<$z$<0.7$; bottom left
  $0.7<$z$<$0.9; bottom right: z$>0.9$.  The symbols are large red
  circles: S/N$\geq$9; large green circles: S/N$\geq$6; medium blue
  circles: S/N$\geq$5; medium black circles: S/N$\geq$4; small black
  circles: S/N$\geq$3 and S/N$\geq$2. The area covered by the T0004
  data release analysed by A10 is shown in magenta.}
\label{fig:xyW1}
\end{figure*}

\begin{figure}
\centering
\mbox{\psfig{figure=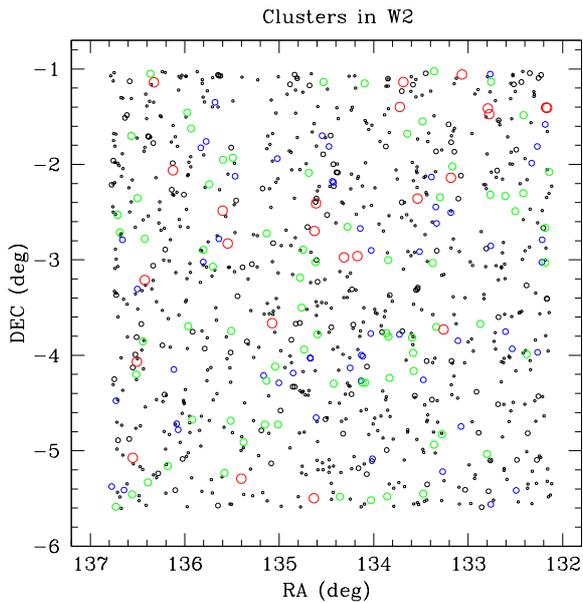,width=8.cm}}
\caption[]{Same as Fig.~\ref{fig:xyW1} for the W2 field. All
    redshifts are included.  }
\label{fig:xyW2}
\end{figure}

\begin{figure}
\centering
\mbox{\psfig{figure=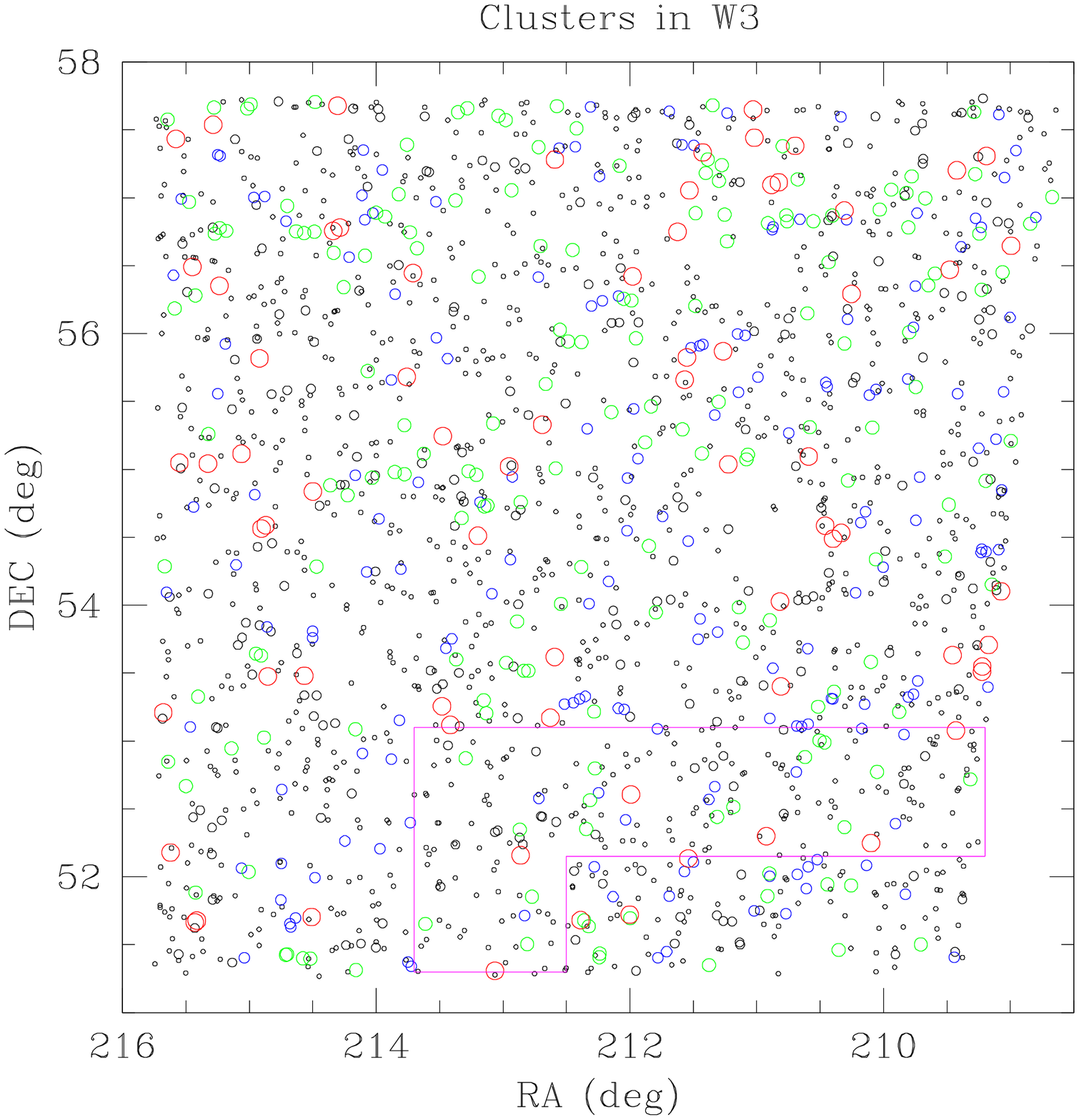,width=8.cm}}
\caption[]{Same as Fig.~\ref{fig:xyW2} for the W3 field.  The area
  covered by the T0004 data release analysed by A10 is shown in
  magenta.}
\label{fig:xyW3}
\end{figure}

\begin{figure}
\centering
\mbox{\psfig{figure=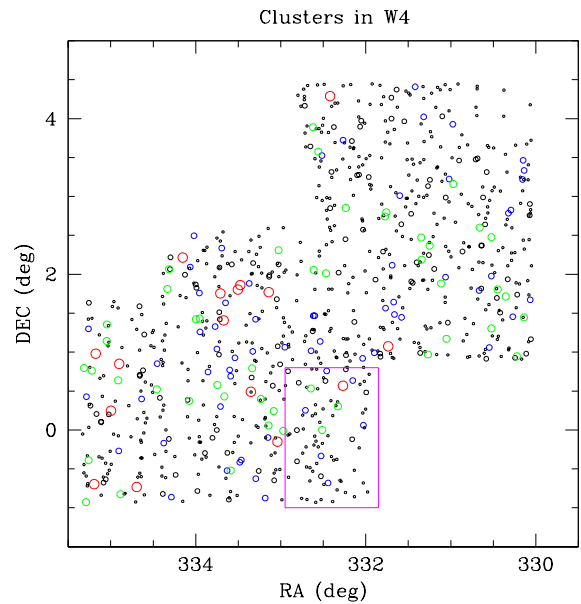,width=8.cm}}
\caption[]{Same as Fig.~\ref{fig:xyW2} for the W4 field.  The area
  covered by the T0004 data release analysed by A10 is shown in
  magenta.}
\label{fig:xyW4}
\end{figure}

We show in Figs.~\ref{fig:xyW1}, \ref{fig:xyW2}, \ref{fig:xyW3}, and
\ref{fig:xyW4} the spatial distributions of our $detections$ in the
four Wide fields (since the W1 field is larger than the other ones,
Fig.~\ref{fig:xyW1} was divided into four redshift bins for the sake
of clarity). As expected, candidate clusters are distributed more or
less homogeneously in all the fields.

\subsection{Redshift distributions of the $detections$}

\begin{figure}
\centering
\mbox{\psfig{figure=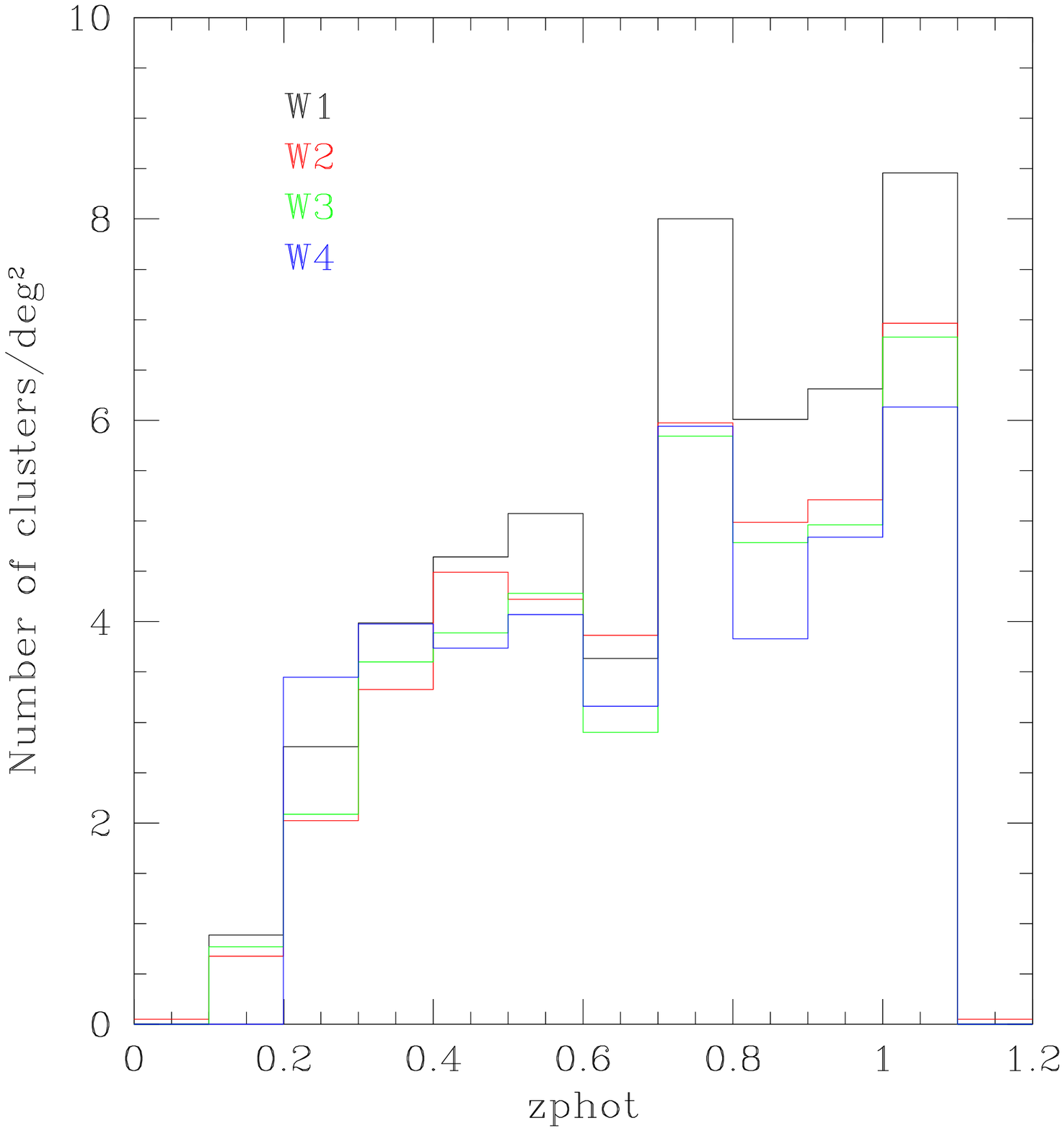,width=7.cm}}
\mbox{\psfig{figure=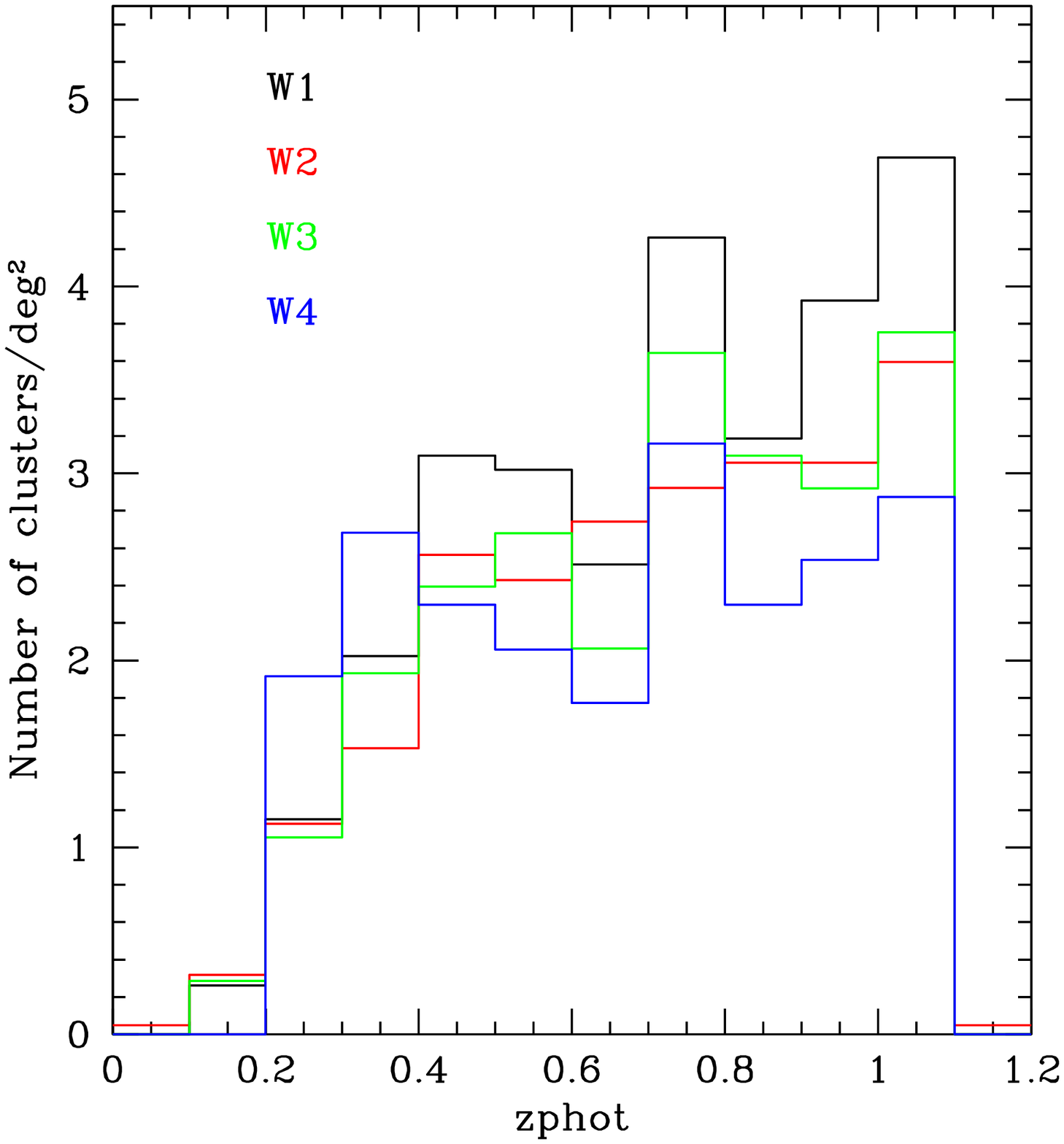,width=7.cm}}
\caption[]{Histogram of the surface density of clusters in photometric
  redshift bins of 0.1 for all the candidate clusters in the four
  CFHTLS-Wide fields (top) and for all the candidates detected at
  3$\sigma _s$ or above (bottom).}
\label{fig:histoz}
\end{figure}

The photometric redshift distributions of our cluster $detections$ in
the four Wide fields are shown in Fig.~\ref{fig:histoz}, where we plot
the density of clusters per square degree in photo-z bins of 0.1
(between $z=0.1$ and 1.15). As expected, these numbers are quite
comparable in the four fields, suggesting that there is no systematic
difference, although the four Wide fields sample very different
regions of the sky.  If we exclude the clusters detected only at the
2$\sigma _s$ level, the shapes of the histograms remain the same, but the
numbers of detected clusters are notably reduced.

\begin{figure}
\centering
\mbox{\psfig{figure=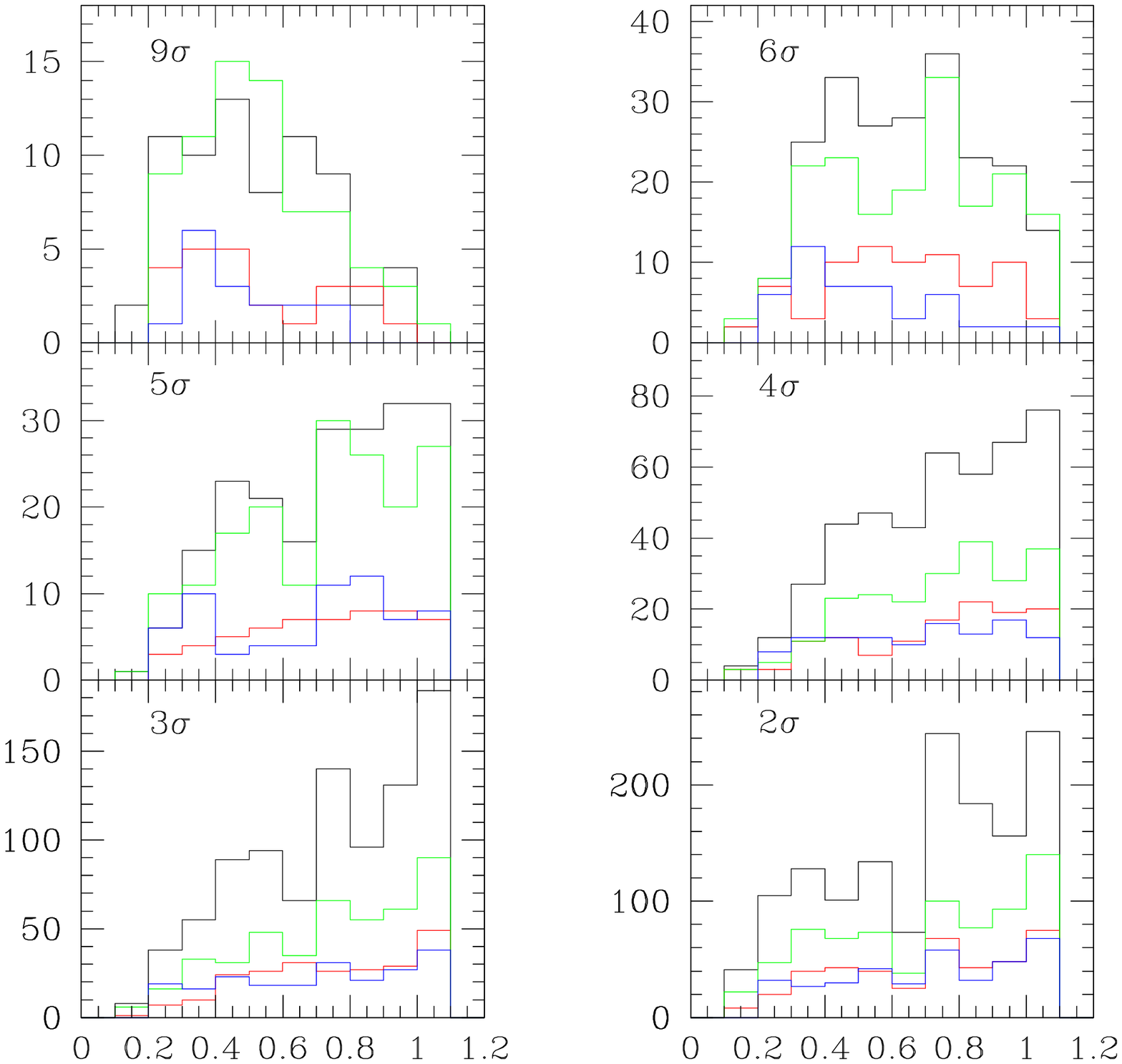,width=8.cm}}
\caption[]{Histograms of the photometric redshift distributions for
  the candidate clusters in the four fields, colour coded as: black
  for Wide~1, red for Wide~2, green for Wide~3 and blue for
  Wide~4. The six figures correspond to the detection levels adopted:
  9$\sigma _s$ (top left), 6$\sigma _s$ (top right), 5$\sigma _s$
  (middle left), 4$\sigma _s$ (middle right), 3$\sigma _s$ (bottom
  left), and 2$\sigma _s$ (bottom right).}
\label{fig:histozcl}
\end{figure}

On the other hand, there seems to be a significant dependence of the
number of $detections$ at high redshift with the signal-to-noise
level. This is illustrated by Fig.~\ref{fig:histozcl}, where we see
that the number of high-redshift clusters detected increases as the
detection level decreases.  We checked that this is consistent with
cosmological predictions.  Evrard et al. (2002) have computed the
expected number of clusters as a function of cluster mass in different
redshift intervals (see their Fig.~7, right column). For clusters of
mass $10^{14}$~M$_\odot$, they predict about four times more clusters
in the redshift range $0.5\leq z<1.2$ than in the range $0.2\leq z<0.5
$. This is consistent with what we find.  Very massive clusters
detected at the 9$\sigma _S$ level are present in small numbers and
are found mainly at redshifts lower than 0.6.

\section{Comparison with other cluster detections in the CFHTLS}
\label{sec:comparison}

Before carrying out analyses based on our catalogue of candidate
clusters, we discuss the validity of the catalogue and its limitations
by comparing our detections with those of other authors.  Several
other cluster candidate catalogues exist in the CFHTLS areas.  The
T0006 release provides better photometric redshifts than previous
versions, but it remains useful to compare our detections with
previous catalogues. In particular, the candidate catalogues of A10
were based on the T0004 release and detection rates were assigned, as
already quoted, by comparisons with the Millennium
simulation. Obviously, one can only estimate the effects of the biases
that were previously included in the simulations. It is therefore also
useful to compare our detections with other real catalogues to uncover
unexpected biaises. Moreover, the modified Millennium simulations
considered in A10 used photometrically-mimicked redshifts based only
on the mean CFHTLS uncertainties and not including the catastrophic
failures sometimes seen in photometric redshifts. We show in the
following that this sometimes has non-negligible effects.

\subsection{Internal assessment}

We  cross-correlated our new catalogues of cluster candidates with
those of A10, matching all candidate clusters located at distances
smaller than 4.2~arcmin (quadratic sum of the typical 3~arcmin
position uncertainty of our candidate clusters) and with photometric
redshifts differing by less than 0.2.  In A10, the total numbers of
candidate clusters were 755, 175 and 99 in the W1, W3, and W4 fields
respectively; the corresponding numbers in the present work are 3247,
1819, and 817. We remind the reader that the solid angle covered is now
notably larger.

Out of the 1029 clusters detected in A10, we redetect only 588
clusters in W1, W3, and W4. To see if the non-redetections occurred in
specific spatial regions, we plotted the positions of the
non-redetected clusters. In all three Wide fields, the non-redetected
clusters are spread over the entire regions covered by A10, so there
is no spatial effect.

\begin{table}
  \caption{Mean values of the  mean $i'-z'$ and $y-z'$ colours in the entire 
    T0006 field, in the entire T0006 cluster sample and in the subcatalogue 
    of galaxies in the 
    clusters detected in T0004 and not redetected in T0006.}
\begin{center}
\begin{tabular}{rrrrrrr}
\hline
\hline
     & T0006   & & T0006 &  & Subcat. &  \\
     & all     & & clusters & & \\
\hline
     & $i'-z'$ & $y-z'$ & $i'-z'$ & $y-z'$ & $i'-z'$ & $y-z'$ \\
\hline
W1   & 0.32 & 0.34 & 0.37 & 0.38 & 0.38 & 0.35 \\
W3   & 0.32 & 0.35 & 0.37 & 0.38 & 0.35 & 0.35 \\
W4   & 0.34 & 0.43 & 0.38 & 0.46 & 0.34 & 0.46 \\
\hline
\end{tabular}
\end{center}
\label{tab:coulmoy}
\end{table}

We also tested the hypothesis that non-redetections could come from
selecting galaxies with $i'\leq 23$ in A10 and galaxies with
$z'\leq22.5$ in the present work could lead to different samples (the
$i'$ filter was replaced by a $y$ filter between the T0004 and T0006
data releases, as explained in Section~2.1).  For this, we compared
the mean $i'-z'$ and $y-z'$ colours in the entire T0006 sample used
here, in the galaxies found to belong to clusters in T0006, and in the
subcatalogues of galaxies in the clusters that were detected by A10
but not redetected in the present work.  The mean values are given in
Table~\ref{tab:coulmoy} for the W1, W3, and W4 fields.  As seen from
this table, only small differences are found, so the different
selections do not seem to introduce a bias.

We also estimated the $i'-z'$ colours in the $i'\leq 23$ and $z'\leq
22.5$ selected samples for the four Wide fields. We find $i'-z'$
between 0.23 and 0.26 for the $i'\leq 23$ sample and $i'-z'$ between
0.32 and 0.34 in the $z'\leq 22.5$ sample. Therefore the fact that we
had to change our magnitude selection between T0004 and T0006
introduces a change in colour of $i'-z'$ smaller than 0.1.

\begin{figure}
\centering
\mbox{\psfig{figure=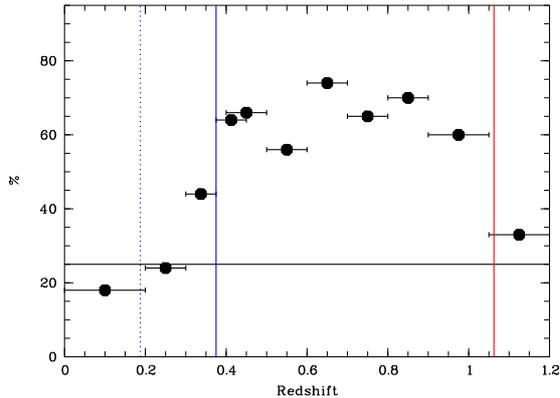,angle=270,width=8cm}}
\caption[]{Percentage of A10 candidate cluster redetection as a
  function of redshift. Horizontal line: statistical random
  probability to have a match between T0004 and T0006 candidate
  clusters. Vertical blue dotted line: redshift where the Balmer break
  passes the middle of the $g'$ band. Vertical blue continuous line:
  redshift where the Balmer break exits the $g'$ band.  Vertical red
  line: redshift where the Balmer break enters the $z'$ band. }
\label{fig:figidentz}
\end{figure}

We now closely examine which T0004 candidate clusters are not
redetected in the T0006 data.  We first test the influence of
redshift. We show in Fig.~\ref{fig:figidentz} the redetection
percentage as a function of the T0004 candidate cluster redshifts. We
see that redetection percentages are close to 70$\%$ between z=0.375
and 1.05. At lower redshifts, redetection percentages are very low,
and not higher than the random probability (computed from the average
number of clusters per unit volume) of a match between T0004 and T0006
candidate clusters (given the number of respective T0004 and T0006
cluster candidates in the considered volume) when considering
redshifts lower than 0.3. This could be explained if the $u^*$ CFHTLS
filter was not very efficient alone in helping to locate the Balmer
break when computing photometric redshifts. It is probably not only by
chance that we see the redetection percentages strongly grow when the
Balmer break exits the $g'$ band, i.e. when photometric redshifts do
not rely only on the $u^*$ band data to locate the break. We find the
same behaviour at redshifts above 1.05, where photometric redshifts
only rely on the $z'$ band to locate the Balmer break.  The redshift
histograms of galaxies in regions of W1, W3, and W4 covered both by
T0004 and T0006 are somewhat different, with more low-redshift
galaxies in T0006 and more high-redshift galaxies in T0004.  Such
differences are due to the improvement in the photometric redshift
determinations between T0004 and T0006.  Fig.~\ref{fig:figidentz}
therefore strongly speaks in favour of only selecting candidate
clusters in the [0.375,1.05] redshift interval.

\begin{figure}
\centering
\mbox{\psfig{figure=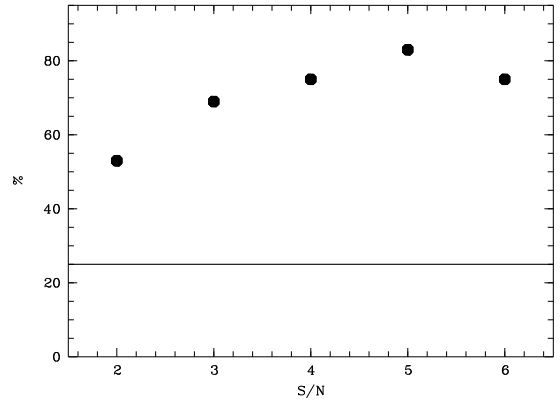,angle=270,width=8cm}}
\caption[]{Percentage of redetections of A10 candidate clusters in the
  [0.375,1.05] redshift interval as a function of candidate
  cluster S/N. The horizontal line is the statistical random probability
  of having a match between T0004 and T0006 candidate clusters. }
\label{fig:figidentsn}
\end{figure}

We now test the effect of the candidate cluster S/N. We show in
Fig.~\ref{fig:figidentsn} the redetection percentage as a function of
the S/N for clusters in the [0.375,1.05] redshift interval where the
redetection percentage is the highest. This figure clearly shows a
regular increase in the percentage from $\sim$55$\%$ when S/N=2 to
$\sim$80$\%$ when S/N$\ge$6.

\begin{figure}
\centering
\mbox{\psfig{figure=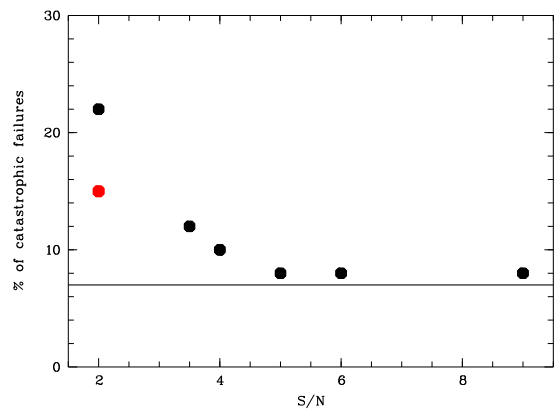,angle=270,width=8cm}}
\caption[]{Black dots: percentage of catastrophic photometric redshift
  failures in the 33 T0006 candidate clusters in the VVDS
  spectroscopic area as a function of S/N. The point at S/N=3.5 was
  obtained by combining the S/N=3 and S/N=4 data.  Red dot: percentage
  of catastrophic photometric redshift failures in the S/N=2 T0006
  candidate clusters in the VVDS spectroscopic area also detected by
  A10. The horizontal line shows the mean percentage of catastrophic
  photometric redshift failures for the whole photometric redshift
  sample (not only in clusters) in the VVDS area.}
\label{fig:cata}
\end{figure}

Low S/N candidate clusters therefore seem to be only poorly
redetected, especially for S/N=2. A possible explanation for this
effect is that part of such poor structures are made of projected
galaxy concentrations on the sky with artificial redshift
concentrations, produced by catastrophic photometric redshift failures
that induce photometric redshift accumulations at given values. Such
features are commonly found, for example, in Coupon et al. (2009), and
are not always easily detected when only a few galaxies are
involved. A way to test this hypothesis is to consider the 33 T0006
candidate clusters in the VVDS spectroscopic area. It is then possible
to compute the percentage of these catastrophic photometric redshift
failures in the corresponding clusters.  A galaxy will be assigned to
a cluster if the redshift difference between the galaxy and the
cluster is smaller than 0.1 and if the galaxy is closer than 1~Mpc (in
projection, calculated at the cluster distance) from the cluster
centre. The same conditions will apply for Section~5. We give in
Fig.~\ref{fig:cata} the percentage of catastrophic photometric
redshift failures in the 33 T0006 candidate clusters in the VVDS
spectroscopic area as a function of the candidate cluster S/N.  The
position of the horizontal line in Fig.~\ref{fig:cata} implies that
the photometric redshifts are always worse for galaxies in clusters
than in the whole field, as already noted by Guennou et
al. (2010). This is true in particular for cluster dominant and bright
spiral galaxies, which are not well fit by the template spectra
available to train photometric redshift codes.  Fig.~\ref{fig:cata}
shows that on the one hand, we clearly see that S/N$\geq$5 candidate
clusters have a behaviour similar to that of the whole sample. On the
other hand, S/N=2 candidate clusters exhibit catastrophic photometric
redshift percentages that are nearly three times higher. However, if
we compute the catastrophic photometric percentages only for those
S/N=2 candidate clusters that were {\it both} detected in A10 with the
T0004 data and with the T0006 data, the percentage value diminishes
strongly. A plot of the photometric versus spectroscopic redshift for
the clusters found in the XMM-LSS survey can be found in Appendix
A.1. of Adami et al. (2011).

  We therefore conclude that S/N=2 candidate clusters detected in A10
  and not redetected with T0006 data are at least partly not
  real. They are probably made of artificial redshift concentrations
  on the sky produced by catastrophic photometric redshift
  failures. This also explains why the S/N=2 candidate cluster
  percentage in Fig.~\ref{fig:figidentsn} (typically 50$\%$) is lower
  than the predicted values in A10 ($\sim$80$\%$). This is because
  catastrophic photometric redshift failures were not taken into
  account when mimicking CFHTLS photometric redshifts with the
  Millennium simulation redshifts. The same applies to the T0006
    sample, but to a lesser extent.

As a conclusion, we therefore typically redetect 75$\%$ of the A10
candidate clusters when considering S/N$\geq$3 structures in the
[0.375,1.05] redshift interval. S/N=2 candidate clusters are also
probably mainly real when detected by {\it both} A10 and the present
paper.

\subsection{Comparison with detections in the Wide fields by other
  authors}

\subsubsection{Comparison with the Thanjavur et al. (2009) detections}

\begin{figure}
\centering
\mbox{\psfig{figure=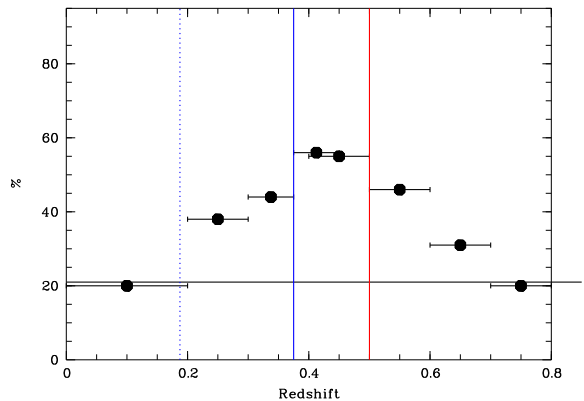,angle=270,width=8cm}}
\caption[]{Percentage of Thanjavur et al. (2009) candidate clusters
  redetected in the present paper as a function of
  redshift. Horizontal line: statistical random probability of a
  match. Vertical blue dotted line: redshift where the Balmer break
  passes the middle of the $g'$ band. Vertical blue continuous line:
  redshift where the Balmer break exits the $g'$ band. Vertical red
  line: redshift where the Thanjavur et al. (2009) method starts
  missing more than 10$\%$ Fornax-like clusters. }
\label{fig:figidentthz}
\end{figure}

\begin{figure}
\centering
\mbox{\psfig{figure=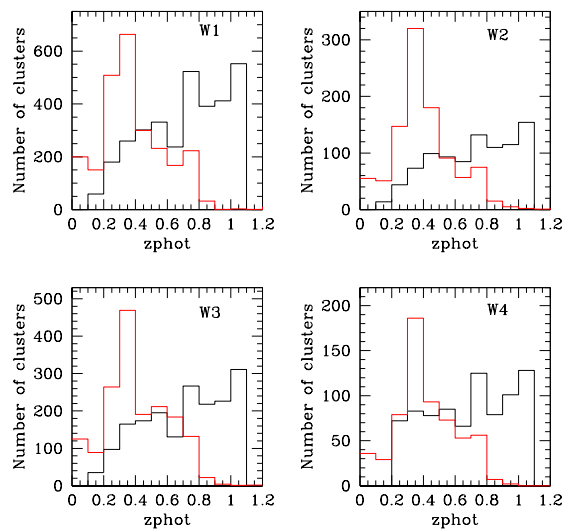,width=7.5cm}}
\caption[]{Histograms of the photometric redshift distributions for
  the candidate clusters detected in the four Wide fields; black: our
  detections, red: detections by Thanjavur et al. (2009).}
\label{fig:histozphotusThan}
\end{figure}

We correlated our catalogues of cluster candidates with those of
Thanjavur et al. (2009) in the four Wide fields. The Thanjavur
catalogues are based on the T0005 CFHTLS data release (which covers
the same field as the T0006 release considered here). They include
2491, 1002, 1697 and 614 galaxy clusters in the W1, W2, W3, and W4
fields, respectively (the respective numbers of clusters in our
catalogues are 3247, 919, 1819, and 817).  We made the same kind of
matching as in the previous section, comparing this time our T0006
cluster candidates with the Thanjavur catalogues.

We first tested the redshift influence. Fig.~\ref{fig:figidentthz} shows
the redetection percentage as a function of redshift. We see that
redetection percentages are close to 55$\%$ between z=0.375 and
0.5. At lower and higher redshifts, redetection percentages are
lower. They are not higher than the random probability of a match
for redshifts lower than 0.2 and higher than 0.7. On the low-redshift
side, the explanation given in the previous section also applies: the
$u^*$ band alone is probably not sufficient to locate the Balmer break
when computing photometric redshifts. On the high-redshift side, we
can note that the Thanjavur method is mostly efficient at finding
clusters at redshifts lower than 0.5 (see
Fig.\ref{fig:histozphotusThan}).  At higher redshifts, Thanjavur et
al. (2009) detect less than 90$\%$ of minor clusters such as Fornax
(see their Figure~3).

\begin{figure}
\centering
\mbox{\psfig{figure=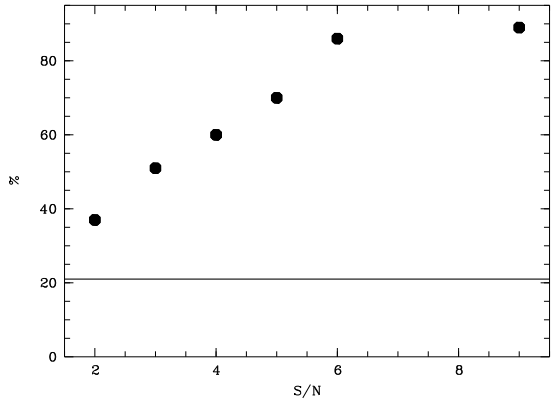,angle=270,width=8cm}}
\caption[]{Percentage of redetections of Thanjavur et al. (2009)
  candidate clusters in the [0.375,0.5] redshift interval as a
  function of the candidate cluster S/N. The horizontal line is the
  statistical random probability of a match. }
\label{fig:figidentthsn}
\end{figure}

We now test the effect of the candidate cluster S/N on the redetection
of the Thanjavur et al. (2009) clusters in the [0.375,0.5] redshift
interval. We show in Fig.~\ref{fig:figidentthsn} the redetection
percentage as a function of the S/N. This figure clearly shows a
regular increase in the percentage from $\sim$35$\%$ when S/N=2 to
$\sim$90$\%$ when S/N$\geq$6. With a stricter condition that the
cluster positions match within 3~arcmin, as taken in Sect.~5, the
percentage increases from $\sim$25$\%$ when S/N=2 to $\sim$75$\%$ when
S/N$\geq$6.

These results are consistent with the conclusions of the previous section.

\subsubsection{Comparison with the Limousin et al. (2009) detections}

Limousin et al. (2009) did a search for (massive) galaxy groups
(i.e. with masses between $10^{13}$ and $10^{14}$~M$_\odot$) in a
large area of 100~deg$^2$ and detected 13 objects in the Wide fields.
These objects are not easy targets for our method, since they are only
groups, therefore not very rich in terms of galaxies. However, we
redetect seven of these 13 groups in our data. Among the six that are
not redetected, three are located in CFHTLS masked regions in at least
one band and therefore are not detectable by our method by
definition. This is because we rely on galaxy detections in the
cluster, while lensing methods are based on peripherical objects and
can detect a cluster even if it is partly in a masked region.  Another
cluster is only detected by one method in Limousin et al. (2009) (at
the edge of a field) and is therefore not very secure. Finally, the
last one is the optically poorest group in Limousin et al. (2009), so
the hardest to detect with our method.  We therefore conclude that we
detect between 80 and 90$\%$ of the Limousin et al.  (2009) groups
accessible by our method.

\subsection{Comparison with detections in the Deep fields}

Obviously, the shallowness of the Wide fields relative to the Deep
fields implies that the numbers of clusters detected in the Wide
fields will be smaller. We have shown in A10 that the ratio of the
numbers of detected clusters in the Deep and Wide fields is expected
to be $2.7 \pm 1.4 $. However, we expect to redetect at least a fraction 
of the clusters with redshifts $z\leq1.15$ found in the Deep fields.

\subsubsection{Comparison with the Olsen et al. (2007) and Grove et
  al. (2009) detections}

Olsen et al. (2007) detect 162 clusters over an area of 3.112~deg$^2$
in the CFHTLS Deep~1, 2, 3, and 4 fields applying the matched filter
method. Out of these, 46 were in Deep~1 and 40 in Deep~3 (Deep~2 and
Deep~4 are not included in W2 and W4, so we cannot try to redetect
their clusters in these fields).  We redetect by our method 20 of
their clusters in W1 and 10 in W3.

Based on the same method, Grove et al. (2009) found 114 clusters, out
of which 34 in the Deep~1 and 27 in the Deep~3 field.  We redetect 20
and 12 clusters. In several cases, we redetect {\it more} clusters
than the numbers found by the matched filter technique. This can be
explained if we assume that we detect many low-mass structures that
are still in the course of their merging process and that will produce
a very massive cluster in the future. In this case, our method will
detect several structures associated with a single candidate cluster
found by the matched filter technique.

The ratio between the considered literature detections and our
redetections is 2.3 (D1) and 4 (D3) for Olsen et al. (2007), and 1.7
(D1) and 2.25 (D3) for Grove et al. (2009). All these numbers are
within the expected ratio of $2.7 \pm 1.4 $.

\subsubsection{Comparison with the XMM-LSS detections}

The most recent cluster catalogue detected by the XMM-LSS was recently
published by Adami et al. (2011), where they present 66
spectroscopically confirmed clusters with redshifts $0.05 \leq z \leq
1.15$ within an area of 6~deg$^2$. Their clusters were divided into
four categories: C0 are clusters detected at optical wavelengths but
not in X-rays, C1 are clusters with the highest probability of
being real clusters, and C2 and C3 are fainter clusters with a decreasing
probability of being truly extended galaxy structures, but which have
passed all the spectroscopic tests.

The number of clusters in the C1, C2, and C3 catalogues of Adami et
al. (2011) and in the area covered by W1 is 40.  By cross correlating
our catalogues with these three catalogues, we redetect 14
clusters. Among the 26 non redetected clusters, 19 are in masked
CFHTLS regions so not detectable by our method. Therefore only 7 are
not detected by our method and should have been. This is
understandable since they are all at redshifts lower than 0.35, where
we have shown that our method is not very efficient.  This represents
a 65$\%$ level detection.

Our detection level depends on the category, but only weakly, and
  with a large dispersion: the detection levels found for the C1, C2,
  and C3 categories for the Adami et al. (2011) clusters are $6.0 \pm
  3.6$, $4.3 \pm 2.1$, and $3.8 \pm 2.7$, respectively.

\subsubsection{Comparison with the Cabanac et al.  (2007), Gavazzi \&
  Soucail (2007) and Bielby et al. (2010) detections}

Cabanac et al.  (2007) detect 42 clusters with a secure redshift in
the four CFHTLS Deep fields, out of which 32 in zones covered by the
Wide fields that we have analysed. We redetect 13 clusters, while 14
of the non redetected clusters are in masked areas. The percentage
of redetection is therefore 68$\%$. 

Gavazzi \& Soucail (2007) detect 14 clusters in the Deep fields, out
of which ten are in the Wide fields. Five are in masked areas and we
redetect two of the remaining five. The last three non-redetected
clusters are all at redshifts lower than 0.16, where our method is not
efficient.

Bielby et al. (2010) searched for clusters with redshifts higher than
1.1 in the Deep~1 and Deep~4 fields. The only cluster at redshift low
enough that we may hope to redetect it is the one at 1.07, but it is
in a partially masked CFHTLS region, which definitively prevents any
detection by our method.

In conclusion, our catalogue of candidate clusters is reliable with
the following limitations: our most reliable detections are between
redshifts 0.375 and 1.05 and for signal to noise$\geq 3$. We now
proceed with a discussion of results derived from our catalogue.

\section{Preliminary analysis of a subsample of stacked clusters}

\begin{figure}[h]
\centering
\mbox{\psfig{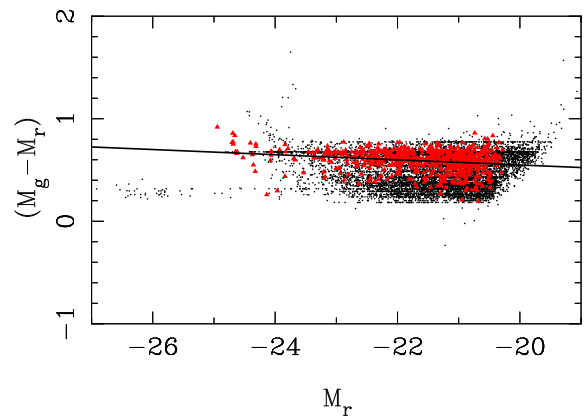}}
\caption[]{Colour-magnitude diagram for 45 stacked clusters (see
  text). The black points are all the galaxies in a radius of 1~Mpc
  around each cluster centre, and with an absolute magnitude
  M$_{z'}<-21.0$. The red points correspond to the subsample of these
  galaxies within $\pm 0.1$ of the cluster redshift.  The black line
  shows the best fit to the colour-magnitude relation drawn from the
  red points}.
\label{fig:cmr45}
\end{figure}

\begin{figure}
\centering
\mbox{\psfig{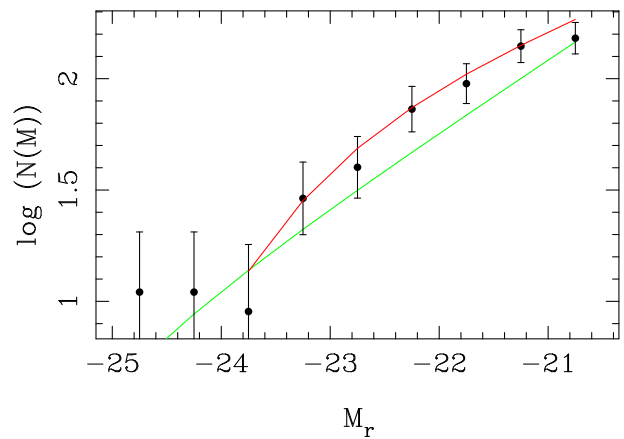}}
\caption[]{Galaxy luminosity function for the galaxies of the 45
  stacked clusters selected in red in Fig.~\ref{fig:cmr45} (showing
  the logarith of the number of galaxies expressed in units of
  galaxies/deg$^2$/0.5~mag). The green line shows the best Schechter
  function fit when all the points are included and the red line shows
  the best Schechter function after excluding the two  brightest
  points.}
\label{fig:glf45}
\end{figure}

As a first test to see how the general properties of the clusters that
we detect match well-known cluster properties, we considered a
subsample of 45 clusters in the W1 field detected simultaneously by
A10, by the present analysis at a significance level of at least
3$\sigma_s$ and by Thanjavur et al. (2009).  We considered the
absolute magnitudes calculated by Le~Phare in order to be able to
stack these clusters, which have redshifts of at most 0.8. For this
redshift, the limiting value $z'=22.5$ chosen for the present analysis
corresponds to an absolute magnitude M$_{z'}=-21.0$. We therefore
applied this absolute magnitude cut to the galaxy catalogue (instead
of the previous magnitude cut at $z'=22.5$), in order to make a galaxy
selection that is independent of redshift. In doing so, we neglected
the k-correction.  Chilingarian et al. (2010) estimate k-corrections
for galaxies of various types between redshifts 0 and 0.5 and an
application is available at http://kcor.sai.msu.ru/. A colour is
requested as input, so we calculated the mean value of $r'-z'$ for all
the galaxies of the W1 field with redshift $0.7<$z$<0.9$ and find
$<r'-z'> = 0.74$. With this value of $r'-z'$, their script gives a
k-correction k$_{z'}$=0.08 in the $z'$ band.  Since this value is
small and quite uncertain, we have decided to apply no k-correction.
We then extracted for each cluster the corresponding galaxy catalogue
within a physical radius of 1~Mpc (projected distance at the cluster
redshift); this radius was computed from the photo-z using the
cosmology calculator by Wright
(2006)\footnote{http://nedwww.ipac.caltech.edu/}. Galaxies were
considered as cluster members if their photo-z was within $\pm 0.1$ of
that of the cluster.

The colour-magnitude diagram for the 45 stacked clusters is shown in
Fig.~\ref{fig:cmr45}. The galaxies considered as belonging cluster
members (the red points in Fig.~\ref{fig:cmr45}) define a 
thinner colour-magnitude relation than that obtained with all the
galaxies located within a projected distance of 1~Mpc (the black
points in Fig.~\ref{fig:cmr45}), as expected if these are indeed {\it
  real} clusters. The linear features appearing at bright magnitudes
are due to saturated objects.  The best fit to the colour-magnitude
relation of cluster galaxies is:
$$(M_{g'} - M_{r'}) = (-0.046\pm 0.099) - (0.025\pm 0.005) \times M_{r'}.$$ 

We have quantified the ``thinness'' of the colour-magnitude
  relation shown in Fig.~\ref{fig:cmr45} by estimating the mean,
  median and standard deviation of the values of $(M_{g'} - M_{r'})$
  for the black and red points of Fig.~\ref{fig:cmr45} in $M_{r'}$
  magnitude bins of 1.  We give these values in
  Table~\ref{tab:coulmag}. In all cases, the dispersions on the values
  of $(M_{g'} - M_{r'})$ are smaller for cluster than for non-cluster
  galaxies, but the differences are too small to be statistically
  significant.

\begin{table}
  \caption{Mean values obtained for the colour-magnitude relation for 
cluster and non-cluster galaxies.}
\begin{center}
\begin{tabular}{rrrrr}
\hline
\hline
Cluster galaxies & & & &  \\
\hline
M$_{r'}$ & M$_{g'}$-M$_{r'}$ & M$_{g'}$-M$_{r'}$ & Nb. of   &   st. dev. \\
        &     mean         &      median     & galaxies &  \\
\hline
$-20.5$ & 0.57 & 0.59 & 178 & 0.12  \\
$-21.5$ & 0.58 & 0.60 & 235 & 0.10  \\
$-22.5$ & 0.61 & 0.63 & 113 & 0.09  \\
$-23.5$ & 0.61 & 0.64 &  38 & 0.10  \\ 
$-24.5$ & 0.68 & 0.68 &  22 & 0.14  \\
\hline
\hline
Non-cluster galaxies & & & &  \\
\hline
$-20.5$  & 0.51 & 0.53 & 3127 & 0.17  \\
$-21.5$  & 0.47 & 0.44 & 3176 & 0.16  \\
$-22.5$  & 0.49 & 0.48 & 1542 & 0.16  \\
$-23.5$  & 0.59 & 0.65 &  478 & 0.16  \\
$-24.5$  & 0.68 & 0.68 &   81 & 0.19  \\
\hline
\end{tabular}
\end{center}
\label{tab:coulmag}
\end{table}

The same galaxies have been used to derive the stacked galaxy
luminosity function (GLF) in the $r'$ band. This GLF is drawn in
Fig.~\ref{fig:glf45}, together with the best fit by a Schechter
function:
$$ S(M) = 0.4 \, \ln 10 \, \phi^{\ast} \, y^{\alpha+1} \, e^{-y} $$
where $y=10^{0.4 \, (M^{\ast}-M)} $. We see from Fig.~\ref{fig:glf45}
that a Schechter function cannot fit the entire GLF, because there is
an excess of galaxies, a feature that is not unusual in clusters (see
e.g. Durret et al. 2011 and references therein).  The best-fit
parameters for the overall fit are $\alpha=-1.90 \pm 0.06$,
M$_{r'}^*-25.9 \pm 0.4$, and $\phi ^*=4.0 \pm 1.6$
(galaxies/deg$^2$/0.5~mag), but this is obviously a bad fit, and the
value found for M$_{r'}^*$ is unrealistically bright.  If we exclude
the two brightest points from the fit, the best-fit parameters become
$\alpha=-1.47 \pm 0.23$, M$_{r'}^*-23.44 \pm 0.47$, $\phi ^*=76 \pm
48$, and the fit is notably better. The faint-end slope is within the
usual range for clusters, i.e. between $\alpha=-1$ and $-2$ (see
e.g. a summary of values of $\alpha$ from the literature given by
Bou\'e et al. 2008 in their Table~A.1).

A full investigation of the properties of the cluster candidates will
be presented in a future paper.

\section{Angular correlation functions}
\label{sec:acf}

As originally shown by Bahcall \& Soneira (1983) and Klypin \& Kopylov (1983),  
the correlation function $\xi_{cc}(r)$ of galaxy clusters can be described 
by a power law:

\begin{equation}
\xi_{cc}(r/r_0)^{-\gamma},
\label{eq:xi3d}
\end{equation}
\noindent
where $\gamma \sim 1.8$, i.e. comparable to the correlation function
of galaxies but with a significantly larger amplitude, depending also
on the richness of the system.

From (\ref{eq:xi3d}) it follows that the angular correlation function is

\begin{equation}
\omega_{cc} = (\theta/\theta_0)^{-\delta}
\end{equation}
\noindent
with $\delta = \gamma - 1 \sim 0.8$.

\begin{figure}[h]
\centering
\mbox{\psfig{figure=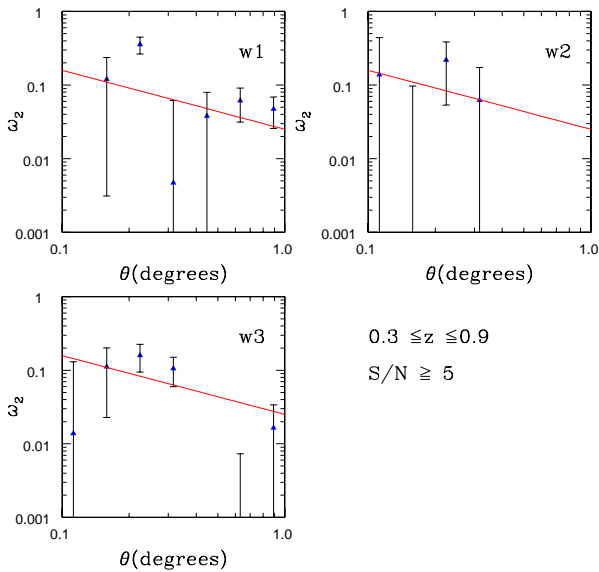,width=8.cm}}
\caption[]{Angular correlation functions for W1, W2, and W3. 
The reference solid line has a slope of $-0.8$ and $\theta_0 = 0.01$
(see text).}
\label{fig:xi2dz03z09sn5}
\end{figure}

\begin{figure}
\centering
\mbox{\psfig{figure=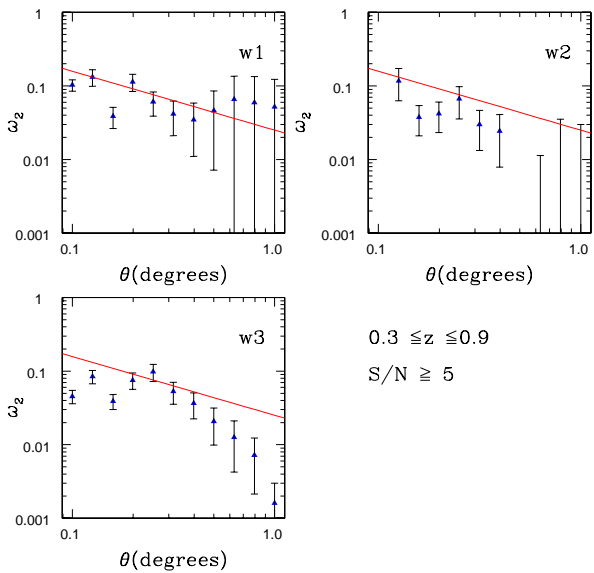,width=8.cm}}
\caption{Area--averaged correlation function for W1, W2, and W3. 
The samples include clusters with $S/N \ge 5$ and $0.3 \le z \le 0.9$.
The reference solid line has a slope of $-0.8$ and $\theta_0 = 0.01$.}
\label{fig:xiave2dz03z09sn5}
\end{figure}

Here we performed a first test to check that the angular correlation
function $w(\theta)$ of the candidate clusters in the Wide fields is
consistent with what has been measured in local cluster samples.  We
assumed that masks are random and do not bias the cluster projected
distribution.  We checked that the best compromise is obtained by
selecting clusters in the redshift range $0.3 \le z \le 0.9$ and
cutting at $S/N \ge 5$ or $S/N \ge 6$. With $S/N \ge 5$ the numbers of
clusters in W1, W2, and W3 are respectively 280, 89, and 229, while
with $S/N \ge 6$ they are 180, 61, and 142.  With a lower $S/N$ cut,
we would have more objects, but we would include a higher fraction of
spurious clusters, poor clusters, and more distant clusters, all
effects that contribute to a lower signal.

To maximize the signal, we chose a large bin in angular separation,
$Log(\Delta\theta) = 0.15$, and present here results obtained by
selecting all candidate clusters with $S/N \ge 5$ in the redshift
range $0.3 \le z \le 0.9$.  We used the Hamilton estimator (Hamilton
1993), which is the best estimator in terms of variance, equivalent to
the Landy--Szalay estimator (Landy \& Szalay 1993; see Labatie et
al. 2011):

\begin{equation}
 \omega(\theta) = \frac{DD(\theta) RR(\theta)}{DR^2(\theta)} - 1
\end{equation}
\noindent
where DD, RR, and DR are the number of cluster-cluster, random-random,
and cluster-random pairs respectively, with angular separation in the
angular bin centred on $\theta$.
Our results are shown in Fig.~\ref{fig:xi2dz03z09sn5}, where as a
reference we plot a solid line with the standard power law
$\omega(\theta) = (\theta/0.01)^{-0.8}$, and points have Poissonian
error bars.

We also estimated the area--averaged correlation function
$\bar{\xi_2}$ from counts in circular cells (see e.g. Cappi \&
Maurogordato 1995). Our results are shown in
Fig.~\ref{fig:xiave2dz03z09sn5}, where errors were derived from
bootstrap resamplings (which underestimate true errors).  As an
integral measurement, the area--averaged correlation function is less
noisy than the direct two--point correlation function.  The data are
consistent with the expected power--law within the large error bars,
and are in rough agreement with previous works on other samples
(included Adami et al. 2010).  A better estimate will require a
selection based on cluster richness and a spectroscopic confirmation
of the candidate clusters.

\section{Large-scale structures around our candidate clusters}
\label{sec:lss} 

\begin{figure}
\centering \mbox{\psfig{figure=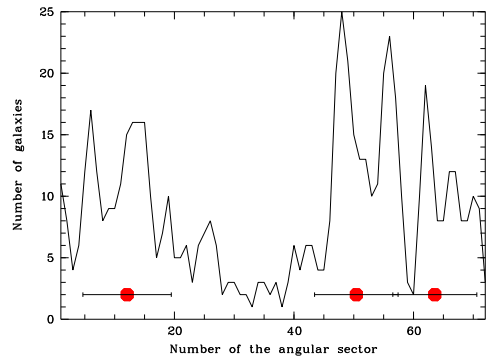,width=7cm,angle=270}}
\caption[]{Significant detected galaxy concentrations (SDGC) around
  one cluster when considering magnitude interval I within a 2.5~Mpc
  radius from the cluster centre (see text). This figure shows the
  number of galaxies in a $\pm 0.05$ photometric redshift bin around
  the considered cluster as a function of the sector number. Filled
  red circles: SDGC detected by SExtractor together with their
  1$\sigma$ level extension. The definition of the angular sector is
  given in the text. }
\label{fig:exampledet}
\end{figure}

\begin{figure}
\centering \mbox{\psfig{figure=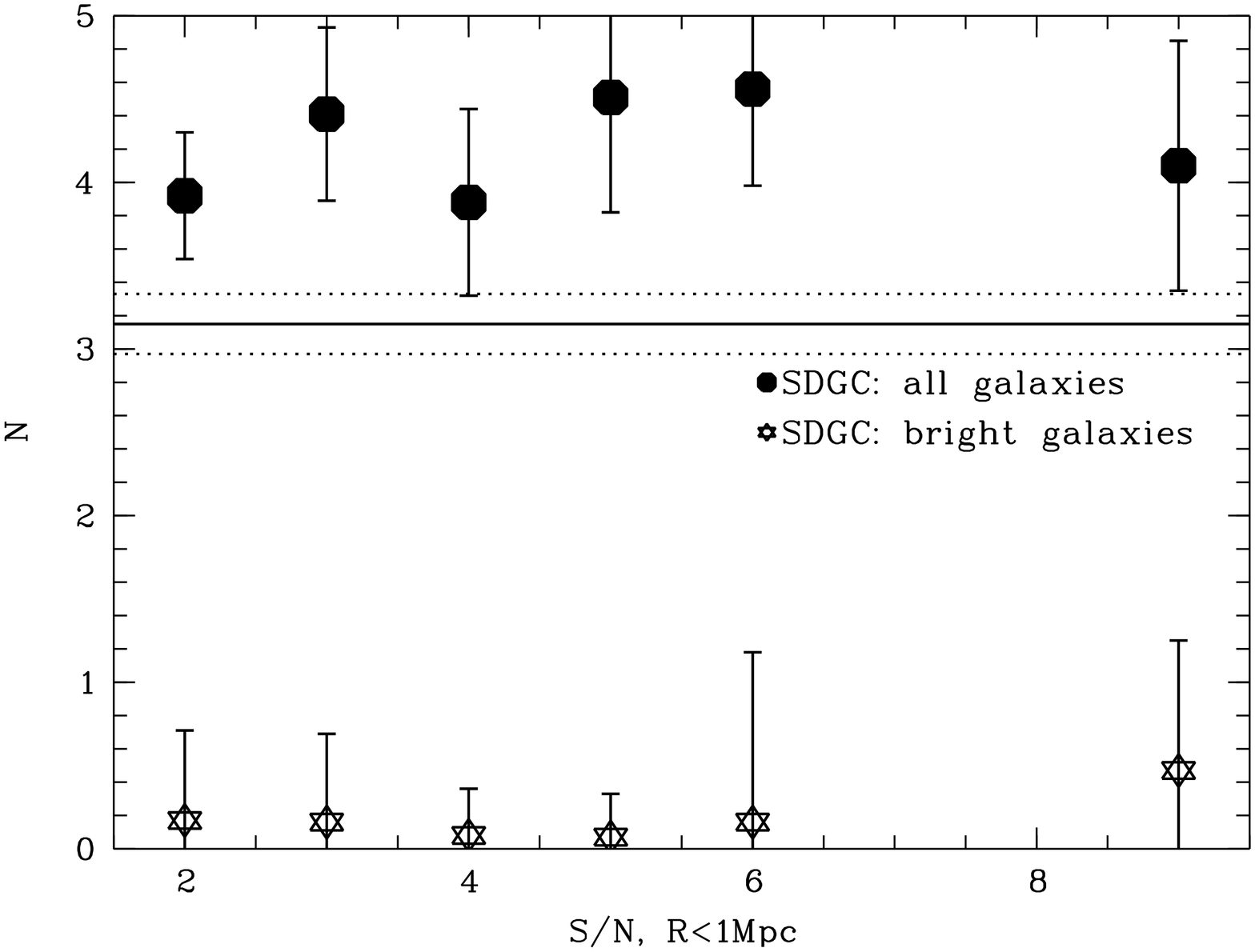,width=7cm,angle=270}}
\centering \mbox{\psfig{figure=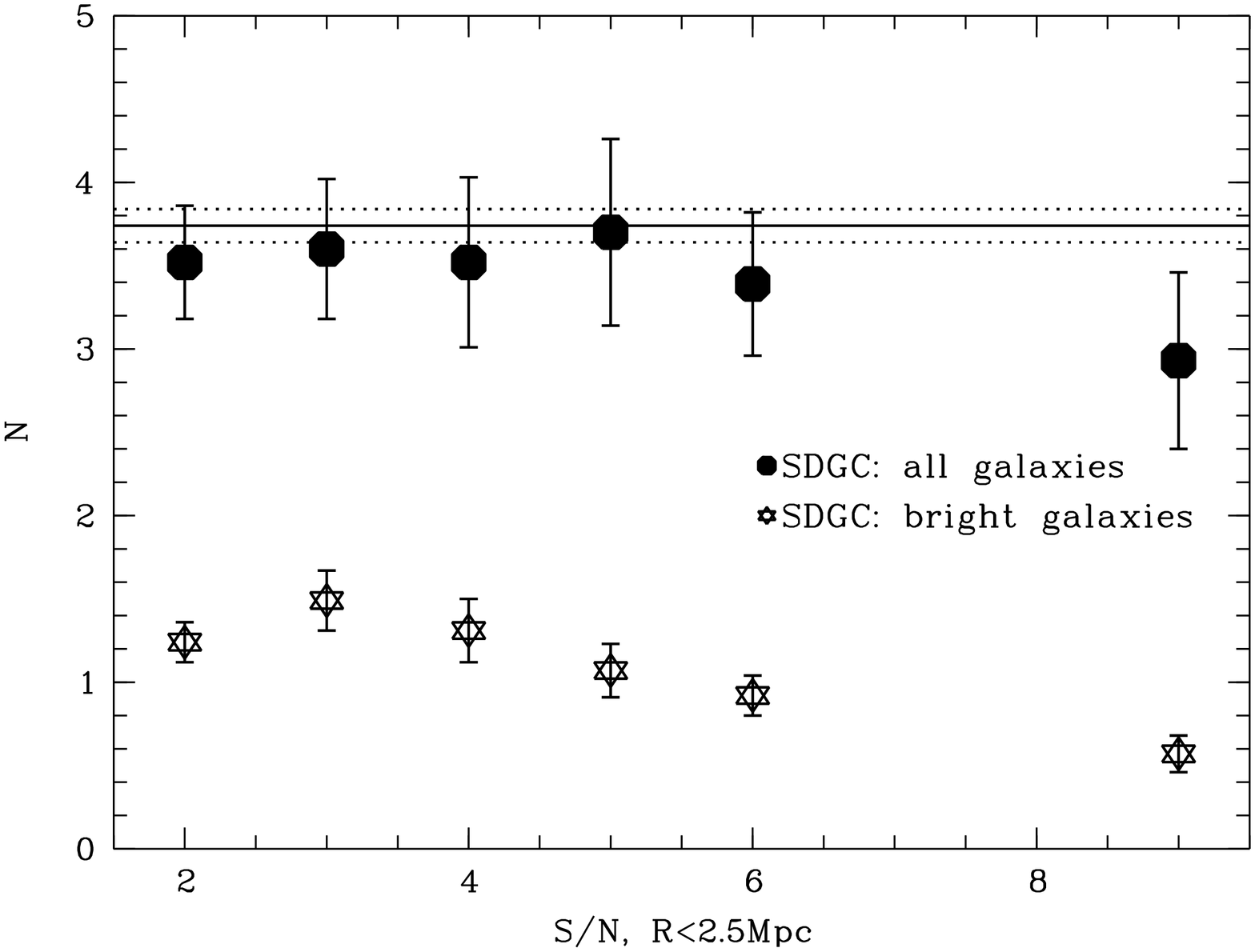,width=7cm,angle=270}}
\centering \mbox{\psfig{figure=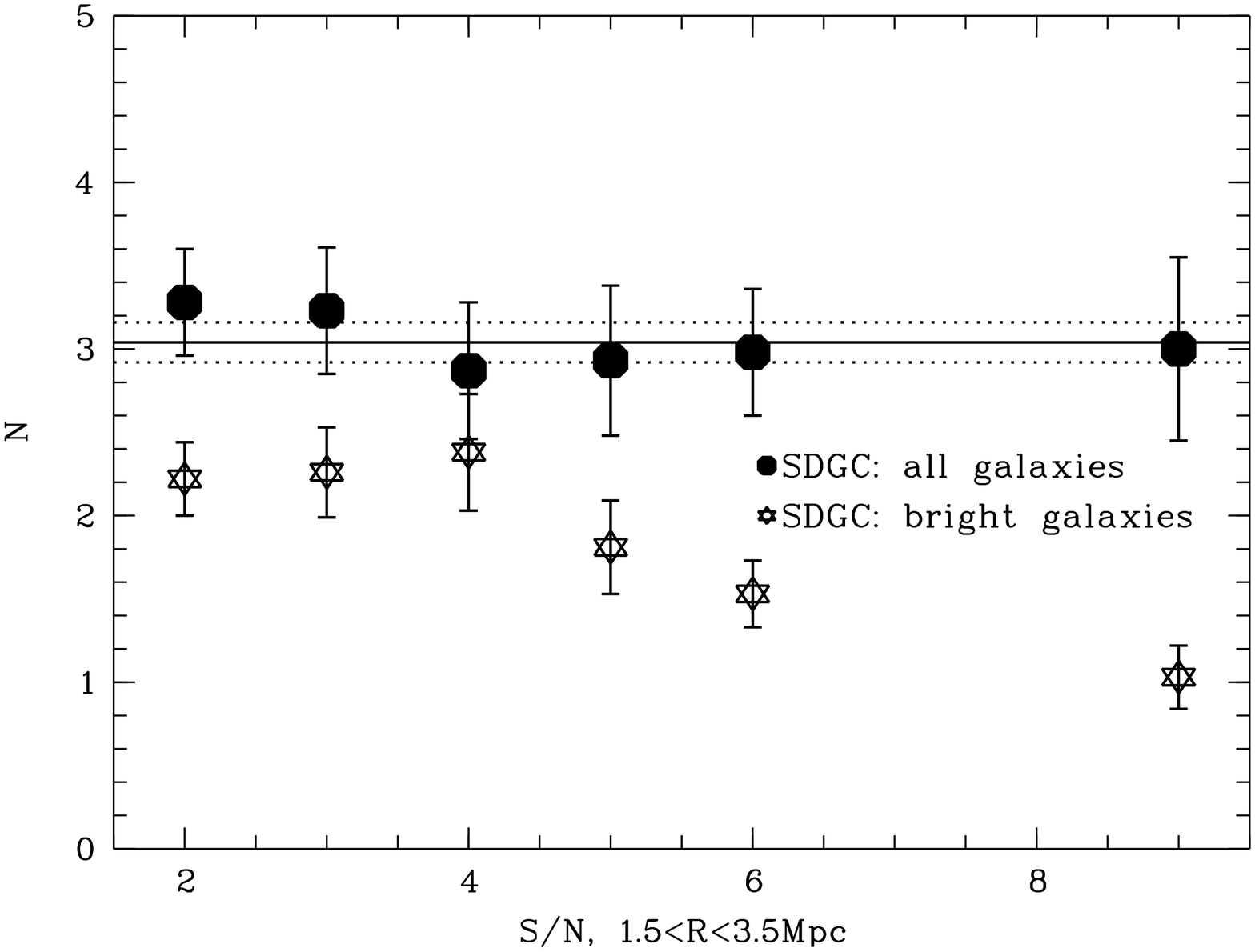,width=7cm,angle=270}}
\centering \mbox{\psfig{figure=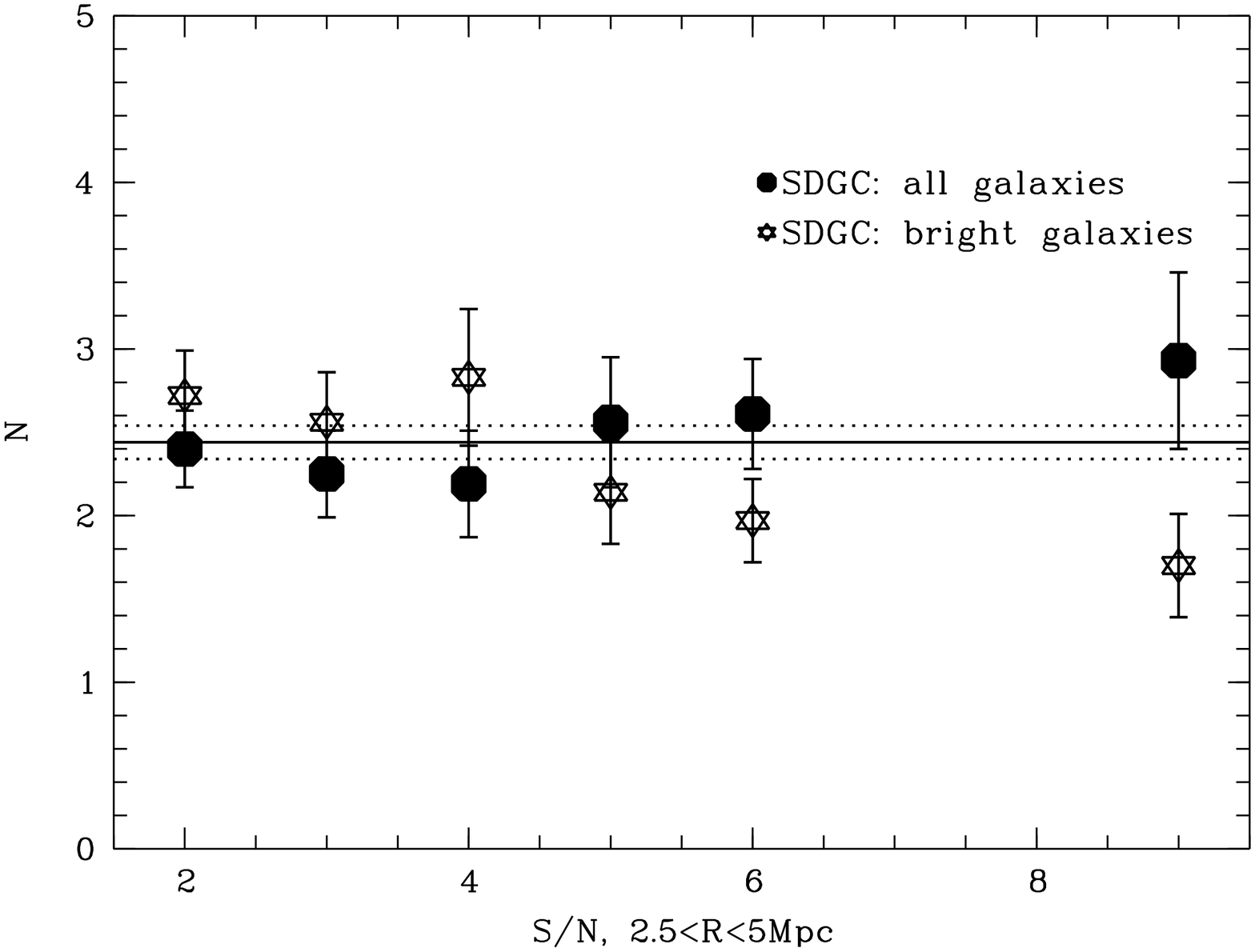,width=7cm,angle=270}}
\caption[]{Mean number of significant detected galaxy concentrations
  (SDGC) around the sample of 491 clusters in W1 from T0006+Th09 as a
  function of cluster detection level (S/N) in the T0006 sample, in
  regions of different radii. From top to bottom the radii are:
  R$<1$~Mpc, R$<2.5$~Mpc, $1.5<$R$<3.5$~Mpc, and
  $2.5<$R$<5$~Mpc. Filled circles: SDGC detected with galaxies in the
  largest possible magnitude interval (see text). Open stars: SDGC
  detected with galaxies brighter than the considered cluster cDs (see
  text). Continuous line: mean number of SDGC in the random sample
  with its error (dotted lines). The uncertainty (interval between the
  two dotted lines) is computed as the second-order momentum of the
  SDGC distribution across the field of view (in pixels of
  0.2$\times$0.2~deg$^2$). By ``all galaxies'' and ``bright
    galaxies'' we mean galaxies belonging to intervals I and II,
    respectively.} 
\label{fig:comparsn}
\end{figure}

\begin{figure}
\centering \mbox{\psfig{figure=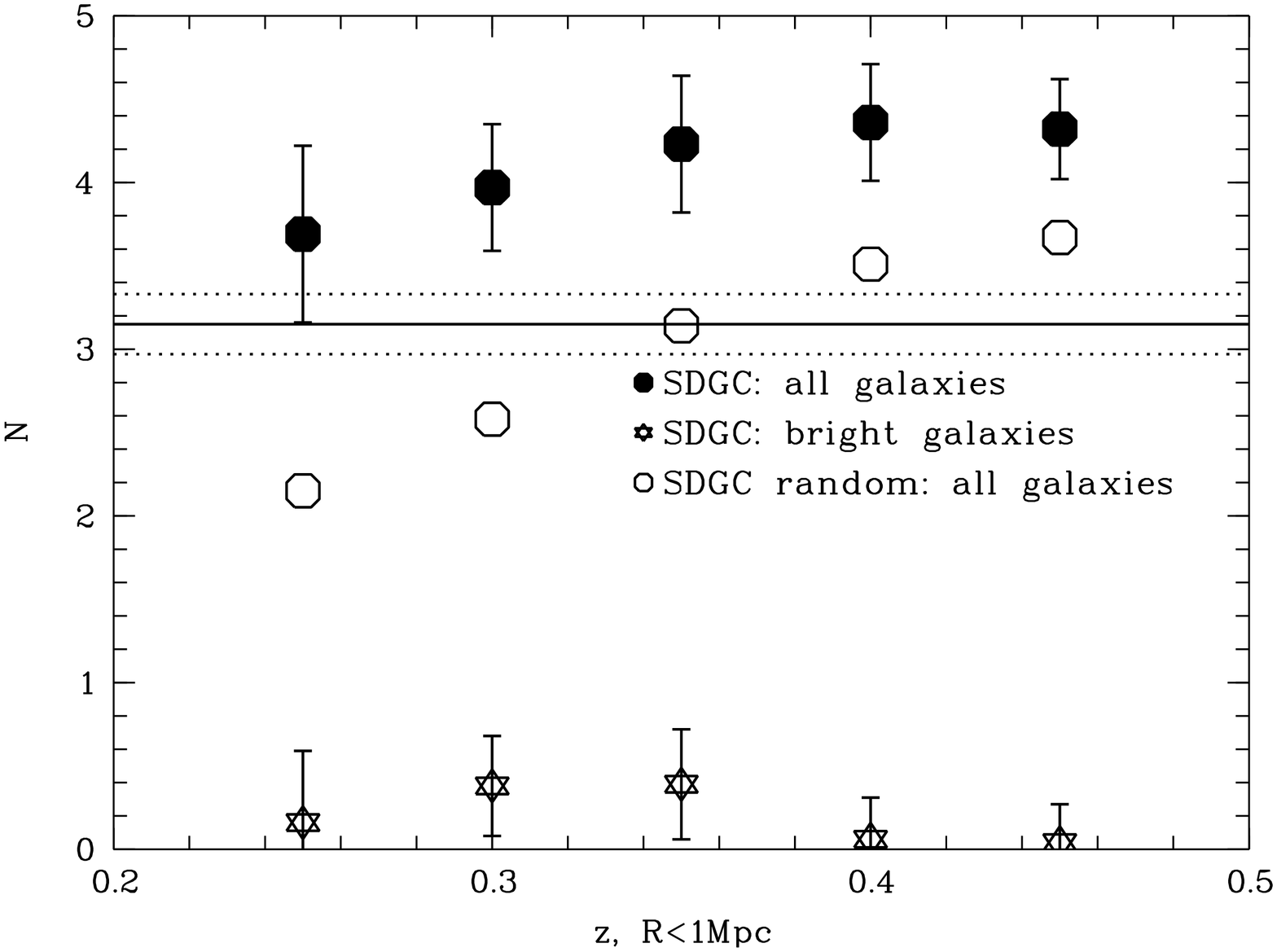,width=7cm,angle=270}}
\centering \mbox{\psfig{figure=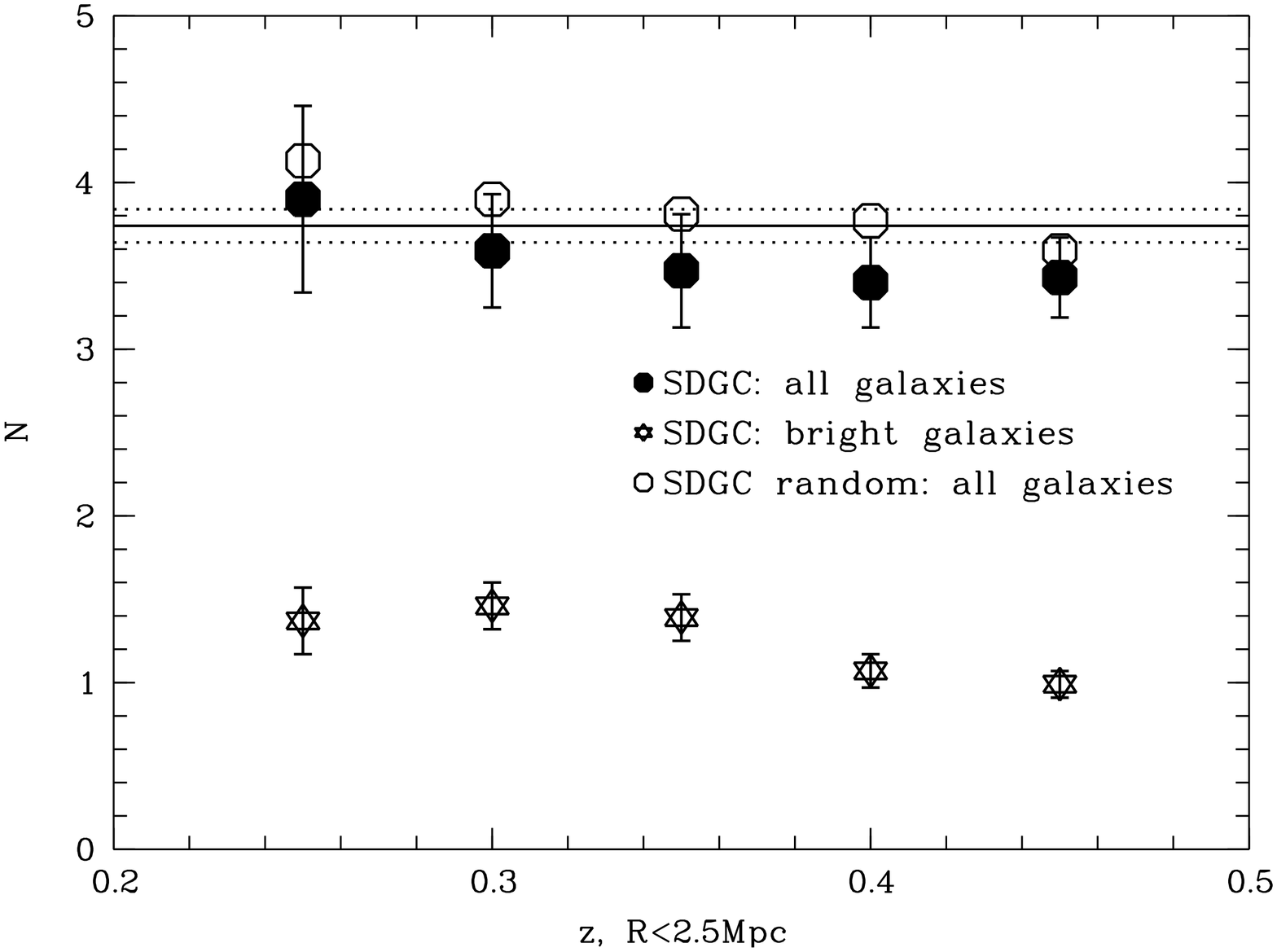,width=7cm,angle=270}}
\centering \mbox{\psfig{figure=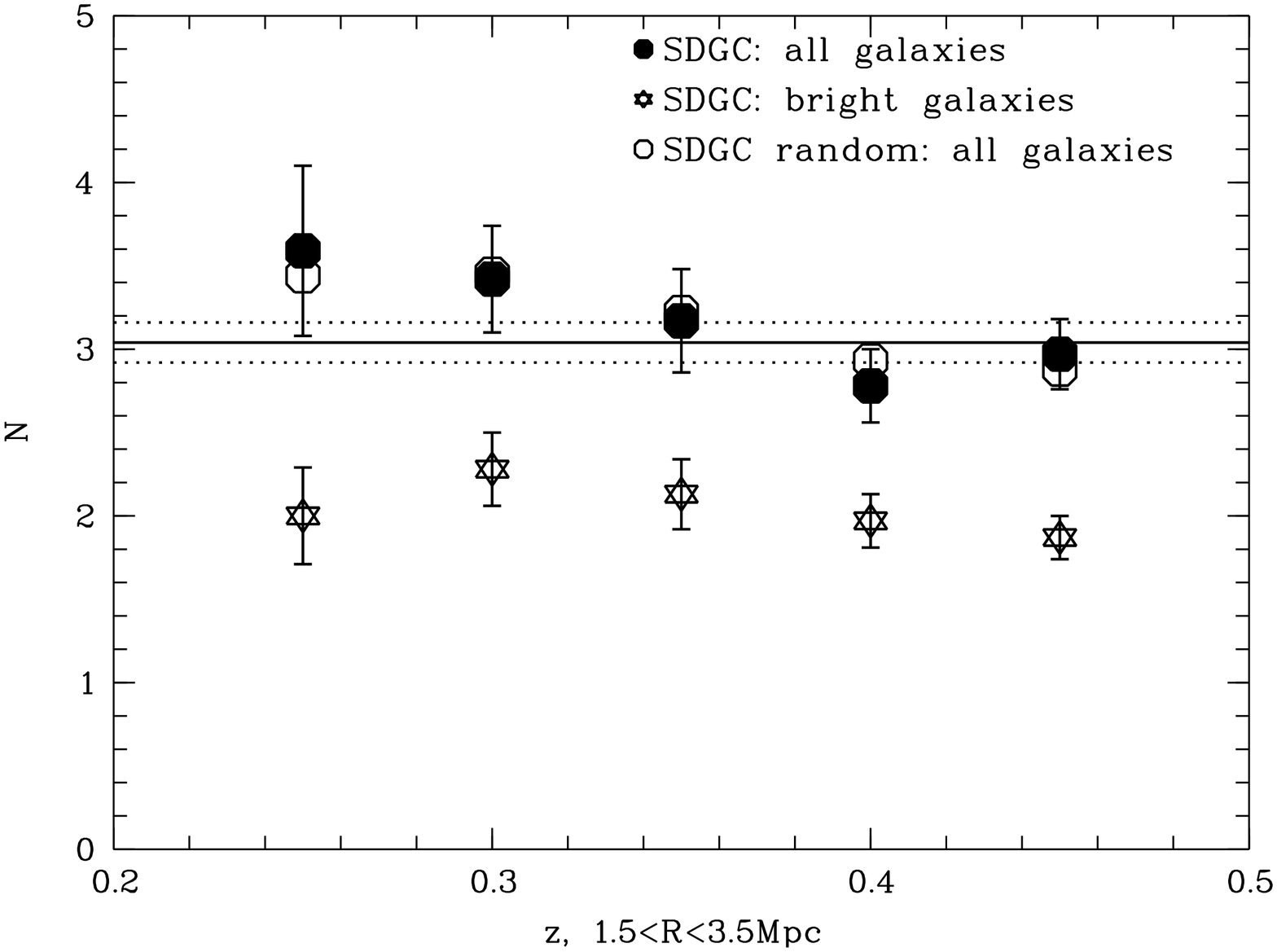,width=7cm,angle=270}}
\centering \mbox{\psfig{figure=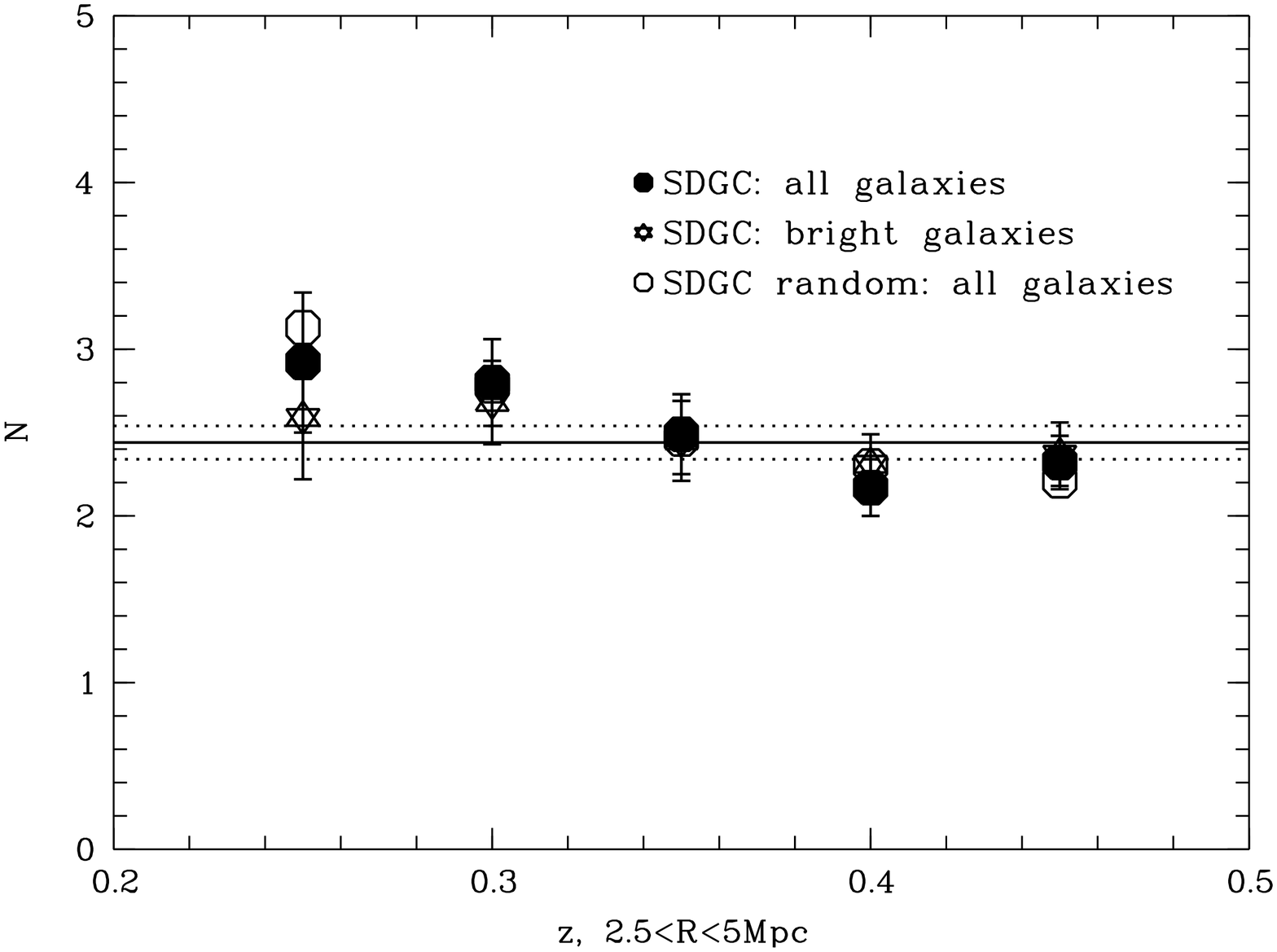,width=7cm,angle=270}}
\caption[]{Same as Fig.~\ref{fig:comparsn} but this time as a
  function of cluster redshift. }
\label{fig:comparz}
\end{figure}

\subsection{Subsample}

The purpose of this section is to broaden the study initiated in Adami
et al. (2011), taking advantage of the very large sample presently
available. However, we cannot use the full sample for several reasons.
First, we demonstrated in Adami et al. (2009) that the CFHTLS Wide
samples are not well suited to detecting large-scale structures
(hereafter LSS) as cosmological filaments at z$\ge$0.5, so we limit
the sample here to z$\leq$0.5 candidate clusters. Second, we need a
precise cluster centre localization in order not to bias the
surrounding LSS detection. The uncertainty on the positions of our
cluster centres is too high for our purposes, so we limit our sample
to clusters also detected by Thanjavur et al. (2009) in the W1 field
(the field with the broadest spatial coverage). Even though clusters
detected by two different methods have a high probability of being
real, the Thanjavur et al. (2009) method provides by definition a good
position of the cluster centre, since it is based on the localization
of the cluster cD.  We present below the LSS analysis of a sample of
491 clusters.

\subsection{Method}

The LSS detection method is close to the one employed in Adami et al.
(2011). Briefly, for a given cluster, we first selected all
galaxies (in a given radius) in a $\pm$0.05 redshift slice around the
cluster redshift. We then counted galaxies in given magnitude intervals
and in 72 angular sectors (5 deg wide and within the considered
radius) around the cluster.

We selected four radii around the clusters. The first one was a 1~Mpc
radius, to consider only the densest cluster areas. The second one was
a 2.5~Mpc radius, to consider areas of the order of the virial
radius. The third one was a corona between 1.5 and 3.5~Mpc to 
consider only galaxy populations close to the virial radius, so just
starting to experience the cluster influence. The last one was a
corona between 2.5 and 5~Mpc to consider galaxy populations not yet
influenced by the considered cluster.

We selected two magnitude intervals. The first one (magnitude
interval~I hereafter) was the full available CFHTLS Wide sample
interval (z'$\leq$22.5) diminished by the distance modulus difference
between a z=0.5 cluster and the considered cluster, corresponding to
the absolute magnitude selection M$_{z'}<-19.75$, chosen to sample the
same absolute magnitude interval for each cluster of the sample. The
second interval we chose (hereafter interval~II) was obtained by
including all galaxies brighter than the cD galaxy of the cluster
+0.5, or M$_{z'}<$M$_{z',cD}+0.5$, to consider only bright galaxies,
and therefore to remove from the detections the LSS consisting only of
faint galaxies. Such faint galaxy populated LSS were, for example,
detected in Adami et al. (2009) around the Coma cluster.

The final step consisted in detecting the significant peaks (assumed
to be LSS or other galaxy groups/clusters) in plots showing the
numbers of galaxies versus the sector angle (see
Fig.~\ref{fig:exampledet}).  In Adami et al. (2011), we chose to
consider a peak as significant if it was at more than three times the
1$\sigma$ variation from the mean value. The 1$\sigma$ level was
computed in this paper by selecting the regions by hand where no peak
was visible. In the study presented here, we used a fully automated
method, better suited to the large number of clusters to analyse. We
ran the SExtractor software on plots of the number of galaxies versus
the one-dimensional sector angle. The deblending threshold was set to 1,
in order to favour the detection of major LSS extending over several
angular sectors. The detection and analysis thresholds were set to~2.

This procedure provided us with a number of significant detected
galaxy concentrations (SDGC hereafter) around the T0006+Th09 sample of
491 clusters in the W1 field. We show an example in
Fig.~\ref{fig:exampledet} of the SDGCs within a 2.5~Mpc radius,
detected for magnitude interval~I.

\subsection{Reference sample}

We also defined a reference sample (i.e. galaxies that are not in
clusters), in order to estimate the statistical behaviour of a random
selection. We selected 491 random positions (both in RA, DEC and in
redshift) in the same area as the real clusters and taken in
interval~I (see above). We then applied the same previous analysis and
defined SDGCs around these random positions. We did not assign S/N
detections or cD galaxy magnitudes to these random positions, so
  of course only interval I can be compared to the random sample.

This allowed us to compute typical uncertainties on the SDGC numbers,
computed as the second-order momenta of the SDGC distributions in the
random sample computed across the field of view (in pixels of
0.2$\times$0.2~deg$^2$).

\subsection{Results}

The results of our analyses are shown in Figs.~\ref{fig:comparsn} and 
\ref{fig:comparz} (with error bars computed as standard deviations).

First, we can see that the mean number of SDGC decreases as a function
of distance to the cluster centre for the interval~I sample (upper to
lower figures in Figs.~\ref{fig:comparsn} and \ref{fig:comparz}). This
result can be interpreted as a mean detection of numerous infalling
structures at radius smaller than 1~Mpc (the densest parts of the
clusters). This is qualitatively consistent with the picture of a
cluster drawn by the Millennium simulations (Springel et
al. 2005). More distant areas would provide direct detections of
cosmological filaments, in particular beyond the virial radius, but
nothing significant is apparent in these areas. Our analyses at radii
smaller than 1~Mpc  most of the time provide at least four directions
with galaxy overdensities (see Figs.~\ref{fig:comparsn} and
\ref{fig:comparz}), suggesting infall rather than just cluster
asymmetries. Moreover, we detect significantly more galaxy
overdensities at radii smaller than 1~Mpc in the real samples compared
to the random samples. This also speaks in favour of a real trend.

Second, there is no strong variation in the mean SDGC value as a
function of the cluster S/N detection when we consider all the
available magnitudes (sample I). However, high S/N clusters (typically
detections greater than 4$\sigma_s$) exhibit lower mean SDGC values,
in particular for the 1.0-3.5~Mpc corona, when considering the
magnitude interval II (only bright galaxies). Clusters therefore seem
to be statistically fed by a comparable number of filaments regardless
of their richness. However, SDGCs populated with bright galaxies are
less numerous beyond radii of 1~Mpc in projection for rich
clusters. This can be interpreted in a simple statistical way: it is
not very likely that a rich cluster has a richer structure in its
direct vicinity, while a poor cluster is more likely to have a rich
companion.

Third, for radii smaller than 2.5~Mpc, we hardly detect any SDGC with
bright galaxies (interval II) compared to the numbers detected in
magnitude interval I. In contrast, nearly all SDGC are detected both
in magnitude intervals I and II beyond radii of 2.5~Mpc. This is in
good agreement with a general depopulation of bright galaxy structures
in the immediate cluster vicinity, as already suggested by Adami et
al. (2011).

Fourth, we do not detect strong variations in the mean SDGC values as
a function of redshift. Apparent variations seen in
Fig.~\ref{fig:comparz} are also visible for the random reference
sample and are therefore probably coming from residual selection
effects.

Fifth, we detect more SDGC at radii smaller than 1~Mpc in the cluster
sample than in the random reference sample. This behaviour is expected
if our cluster sample really includes massive structures formed at the
nodes of the cosmological filaments. Conversely, we detect slightly
fewer SDGCs for radii smaller than 2.5~Mpc in the cluster sample than
in the random reference sample. This is also expected if, as
previously suggested, clusters are depopulating their vicinity just
beyond their densest areas. Finally, when considering areas at radii
larger than the typical virial values, we have exactly the same mean
SDGC values when considering the cluster or the random reference
sample. This tendency is detected both as a function of the cluster
detection S/N and redshift. One more time, this is expected as we are
dealing with areas where the clusters no longer have any influence.

\section{Discussion and conclusions}
\label{sec:concl}

We have searched for clusters of galaxies in the full CFHTLS Wide
survey, based on the T0006 data release. Our method is based on the
galaxy photometric redshifts computed with Le Phare, limited to
magnitudes $z' \leq 22.5$. We constructed galaxy density maps in
photometric redshift bins of 0.1 based on an adaptive kernel
technique, detected structures with SExtractor at various detection
levels, and built cluster catalogues by applying a minimal spanning
tree algorithm.  We detected several thousand candidate clusters with
estimated mean masses between 1.3 10$^{14}$ and 12.6 10$^{14}$
M$_\odot$, thus increasing the number of known optical high-redshift
cluster candidates by a large factor, an important step towards
obtaining reliable cluster counts to measure cosmological parameters.

The cross-identification of the clusters in our catalogues with
various other cluster searches in these regions based on other
methods gives consistent results, in particular for clusters that we
detect at 3$\sigma$ or more.
By stacking a subsample of 45 clusters, we show that this stacked
cluster indeed has ``normal'' cluster characteristics
(colour-magnitude relation, galaxy luminosity function).
The cluster-cluster correlation function is comparable to the one
obtained in A10 (slope and amplitude). 

We also analysed galaxy distributions in cosmological filaments, based
on a sample of 491 clusters detected both by our method and by
Thanjavur et al. (2009) in the W1 field. For these clusters, we
counted the numbers of galaxies in 72 sector angles around the cluster
and searched for significant peaks in these distributions. This
procedure provided us with a number of significant detected galaxy
concentrations (SDGC hereafter).  We find that the mean number of SDGC
decreases as a function of distance to the cluster centre, as expected
if we are detecting infalling structures at projected radii smaller
than 1~Mpc (the densest parts of the clusters).  There is no strong
variation in the mean SDGC value as a function of the SDGC S/N
detection, though high S/N clusters tend to exhibit lower mean SDGC
values. Clusters therefore seem to be statistically fed by a
comparable number of filaments, regardless of their richness. However,
SDGCs populated by bright galaxies are less numerous beyond radii of
1~Mpc for rich clusters.  We find a general depopulation of bright
galaxy structures in the immediate cluster vicinity, as already found
by Adami et al. (2011).  We do not detect strong variations of the
mean SDGC values as a function of redshift.  We detected more SDGC at
radii smaller than 1~Mpc in the cluster sample than in the random
reference sample and slightly fewer SDGCs for radii smaller than
2.5~Mpc, as expected if our cluster sample really includes massive
structures formed at the nodes of the cosmological filaments, while
clusters are depopulating their vicinity just beyond their densest
areas.

The 3D spatial distributions of the candidate clusters detected in all
the Wide fields show no obvious concentration of clusters or
large-scale structures. In view of all the properties presented in
this paper, our detected clusters behave as expected if they are
located at the intersection of filaments by which they are fed.

A more detailed analysis of the properties of this sample of clusters
and of the associated cosmological implications is beyond the scope of
the present paper and will be done in a future work.

\begin{acknowledgements}

  We acknowledge the referee's very thorough reading and constructive
  comments, which helped to improve the paper.  I.M. acknowledges
  financial support from the Spanish grant AYA2010-15169 and Junta de
  Andaluc\'{\i}a TIC114 and the Excellence Project P08-TIC-03531.  We
  are grateful to K. Thanjavur for sending us his catalogues prior to
  publication.

The Millennium Simulation databases used in this paper and the web
application providing online access to them were constructed as part
of the activities of the German Astrophysical Virtual Observatory.

\end{acknowledgements}

\appendix

\section{Galaxy selection}

\begin{center}
\begin{table*}
  \caption{Number of galaxies in the four Wide fields.  }
\begin{center}
\begin{tabular}{lllll}
\hline
\hline
      & Wide 1 & Wide 2 & Wide 3 & Wide 4 \\
\hline
Total number (N)                   &5,346,671&2,100,453&3,858,738&2,040,682 \\
N($0<z'\leq 22.5$)                 &2,344,677&1,031,367&1,673,192&1,060,507 \\
N($0<i'\leq 23$ or $0<y \leq 23$)    &2,658,897&1,137,413&1,900,186&1,162,730 \\
\hline
\end{tabular}
\end{center}
\label{tab:ngal}
Note: second line: galaxies with magnitudes $0<z'\leq 22.5$, third line: galaxies with magnitudes $0<i'\leq 23$ or $0<y\leq 23$.
\end{table*}
\end{center}

The numbers of galaxies in the various catalogues used in this paper
are given in Table~A.1. We can see that the selection criterium based
on the $z'$ magnitude chosen here ($z'\leq 22.5$) gives numbers of
galaxies comparable to those obtained with the criterium based on $i'$
magnitudes by A10 ($0<i'\leq 23$), confirming that it is meaningful to
compare the present results directly with those of A10. However, the
$z'\leq 22.5$ selection tends to make us lose low-redshift galaxies
and gain few high-redshift galaxies, as discussed in Section~2.1.

\end{document}